\pdfoutput=1  
\documentclass[aps,prd,superscriptaddress,floatfix,nofootinbib,preprintnumbers,
eqsecnum,twocolumn,longbibliography
]{revtex4-1}

\usepackage{amsmath,float,graphicx,placeins,amssymb,multirow,pgfplots,pgfplotstable}
\usepackage{tikz}
\usepackage{soul}
\setstcolor{red}
\usepackage{braket}
\usepackage{booktabs}

\usepackage{lipsum} 
\usepackage{comment}
\usepackage{enumerate,slashed,cancel,comment,color,bbm,graphicx,rotating,placeins,mathtools,multirow,hhline}
\usepackage[normalem]{ulem}
\usepackage{array}
\usepackage{enumitem}
\usepackage{hyperref}
\usepackage{color}
 \hypersetup
 { 
   colorlinks,
   linkcolor={blue!80!black},
   citecolor={blue!70!black},
   urlcolor={blue!70!black}
 }

\usepackage{diagbox}

\usepackage{subfig}
\usepackage{caption}

\usetikzlibrary{shapes.geometric,arrows}
\graphicspath{}

\begin{document}

\title{Quantum Annealing: Optimisation, Sampling, and Many-Body Dynamics}

\def\andname{\hspace*{-0.5em}} % gets rid of "and" in author list
\author{Steven Abel}
\email[Email address: ]{s.a.abel@durham.ac.uk}
\affiliation{IPPP and Department of Mathematical Sciences, Durham University, Durham, DH1 3LE, United Kingdom}
\author{Andrei Constantin}
\email[Email address: ]{a.constantin@bham.ac.uk}
\affiliation{School of Mathematics, University of Birmingham, Watson Building, Edgbaston, Birmingham B15 2TT, United Kingdom}
\affiliation{Rudolf Peierls Centre for Theoretical Physics, University of Oxford, Parks Road, Oxford OX1 3PU, United Kingdom}
\author{Luca A. Nutricati}
\email[Email address: ]{luca.nutricati@physics.ox.ac.uk}
\affiliation{Rudolf Peierls Centre for Theoretical Physics, University of Oxford, Parks Road, Oxford OX1 3PU, United Kingdom}

\begin{abstract}
\noindent
{\bfseries Abstract.} 
Quantum annealing is a computational paradigm in which optimisation problems are mapped onto the energy landscape of an interacting quantum system and explored through its dynamical evolution. By continuously transforming a simple initial Hamiltonian into one whose ground state encodes the solution, the system traverses a complex landscape via a combination of quantum fluctuations, tunnelling processes, and dissipative dynamics. Unlike gate-based quantum computing, quantum annealing is a specialised and near-term approach aimed primarily at discrete optimisation and sampling tasks. While it is not expected to provide polynomial-time solutions to NP-hard problems in the worst case, it offers a physically motivated heuristic for navigating rugged energy landscapes that arise across science and engineering. Modern quantum annealers realise programmable spin systems with thousands of qubits, placing them among the largest controllable quantum devices currently available. As a result, their significance extends beyond optimisation: they also function as experimental platforms for studying non-equilibrium many-body quantum dynamics in regimes that are difficult to access using classical simulation.

In this review we present an accessible introduction to the principles of quantum annealing, describe the main hardware platforms and algorithmic techniques, and analyse how tunnelling, spectral gaps, and open-system effects shape computational performance. We survey applications ranging from optimisation and machine learning to quantum simulation and many-body physics, and discuss the central challenges in benchmarking, scaling, and control. These perspectives position quantum annealing as a distinctive framework at the interface of optimisation, stochastic sampling, and programmable quantum dynamics, with a role that is complementary to both classical algorithms and gate-based quantum computing.

\end{abstract}
\maketitle

\vspace{12pt}
\tableofcontents

\def\beq{\begin{equation}}
\def\eeq{\end{equation}}

\section{Introduction}
\label{sec:qa_intro}

\subsection{Quantum Annealing as a Computational and Physical Framework}

Quantum annealing (QA) occupies a distinctive position within the modern landscape of quantum technologies. Originally proposed as a heuristic method for solving combinatorial optimisation problems through adiabatic quantum evolution, it has developed into a broader framework encompassing optimisation, sampling, analogue quantum simulation, and the study of many-body quantum dynamics. Modern quantum annealers realise programmable interacting spin systems containing thousands of qubits, making them not only computational devices but also experimental platforms for exploring complex quantum behaviour at scales that are difficult to access using classical simulation.

Unlike gate-based quantum computing, which aims for universal digital control of quantum information, QA is based on continuous-time Hamiltonian evolution. In this paradigm, a quantum system is steered from a simple initial Hamiltonian to one that encodes a problem of interest, and the computation is realised through the resulting many-body dynamics. Here, \emph{quantum dynamics} refers to the time evolution of a quantum system under its governing Hamiltonian, which determines how its state changes through coherent interactions and external control. In the ideal adiabatic limit, this evolution is approximately coherent and isolated, and the approach connects directly to universal quantum computation. In practice, however, present-day devices operate in regimes where quantum coherence, thermal effects, and control imperfections all play a role. As a result, QA is best understood not as a single algorithm, but as a \emph{physical process} whose computational behaviour emerges from the interplay between coherent quantum dynamics, environmental interactions, and problem structure.

This dual character places QA at an unusual intersection between physics and computation. On the one hand, it provides a programmable framework for encoding discrete optimisation problems into Ising or QUBO Hamiltonians (introduced in more detail below). On the other hand, it offers access to large-scale, controllable quantum many-body dynamics that are difficult to simulate classically. These perspectives are closely intertwined: the same device that searches for low-energy configurations can also act as a sampler for structured distributions or as a laboratory for studying non-equilibrium quantum phenomena. Understanding when and how these roles become useful is central to assessing the capabilities~of~QA.

At its core, QA is motivated by a simple but far-reaching question: can quantum dynamics provide new ways of exploring complex \emph{energy landscapes}? By this we mean the function that assigns an energy value to each possible configuration of the system, as determined by the problem Hamiltonian. This landscape may be discrete (as in Ising models) or continuous in more general settings, and typically contains many local minima separated by barriers. In optimisation contexts, configurations with lower energy correspond to better solutions, but the same concept also arises more broadly in physics when analysing the structure of many-body systems. From an optimisation perspective, the question is whether quantum effects such as superposition and tunnelling can assist in locating low-energy configurations of difficult combinatorial problems. From a physics perspective, it concerns how many-body quantum systems evolve under time-dependent Hamiltonians and how they traverse competing low-energy configurations. In this sense, QA may be viewed as computation realised through controlled many-body quantum dynamics.

Among its various roles, the most immediate and historically influential motivation for QA comes from optimisation. The key idea is that a wide class of discrete optimisation problems can be reformulated as the search for low-energy states of interacting spin systems.

\subsection{Optimisation problems and Ising encodings}

A wide range of problems in science and engineering can be cast in a common form: we are given a set of possible configurations together with a \emph{cost function} that assigns a numerical value to each configuration, and the goal is to find the configuration of lowest cost. Problems of this type are collectively known as \emph{optimisation problems}. Although such problems are classical in nature, the central idea of QA is to embed them into a quantum system whose dynamics can be controlled. In this mapping, classical configurations correspond to basis states of a quantum Hamiltonian, and the cost function becomes its energy. The resulting quantum evolution provides a physically motivated search process over the configuration space, in which effects such as superposition and tunnelling can connect configurations that may be difficult to access using purely classical dynamics.

In some cases the variables involved are continuous, as in parameter estimation or data fitting. In many important situations, however, the variables are discrete: possible routes and candidate timetables in a transportation network, spin assignments in a magnetic material, or truth values in a satisfiability problem. These \emph{combinatorial} optimisation problems frequently become intractable as their size increases, even when using the most powerful classical computers. Continuous problems can often be discretised and treated using similar methods.
A convenient representation for many such problems is provided by the \emph{classical Ising model}, defined in terms of binary variables (``spins'') $s_i = \pm 1$ and an energy function
\begin{equation}
H_{\text{Ising}}(\{s_i\}) = \sum_i h_i s_i + \sum_{i<j} J_{ij} s_i s_j,
\label{eq:ising}
\end{equation}
where $h_i$ and $J_{ij}$ are real parameters. Finding the ground state of this Hamiltonian is itself an optimisation problem, since it corresponds to identifying the configuration of minimum energy.

A closely related formulation is the \emph{Quadratic Unconstrained Binary Optimisation} (QUBO) problem, in which one minimises a quadratic function $x^T Q x$ of binary variables $x_i\in\{0,1\}$. The two formulations are equivalent under the change of variables $s_i = 2x_i - 1$, so that every Ising Hamiltonian corresponds to a QUBO instance, and vice versa. In the QUBO representation, linear terms appear as diagonal entries of $Q$ (since $x_i^2 = x_i$), while pairwise interactions correspond to off-diagonal elements.

The Ising formulation plays a central role in QA for two complementary reasons. First, it provides an expressive encoding framework: a wide class of combinatorial optimisation problems can be reformulated as the task of finding the ground state of an Ising Hamiltonian. In particular, many standard NP-complete and NP-hard problems---classes of problems for which no efficient solution methods are known and whose computational cost is widely believed to grow rapidly (often exponentially) with system size (see Appendix~\ref{sec:ComplexityTheory} for a brief overview)---admit explicit polynomial-time mappings to this form~\cite{Lucas:2013ahy}. Here, ``polynomial-time'' refers to the cost of constructing the Ising or QUBO representation, not to solving the resulting optimisation problem, which remains computationally hard in general. As a consequence, the Ising framework provides a unified representation for a large family of optimisation problems, including many that are NP-hard, for which the associated search spaces typically grow exponentially with system size and no general efficient solution methods are known.

Second, the parameters $h_i$ and $J_{ij}$ correspond directly to experimentally tunable quantities in hardware platforms, currently including superconducting flux qubits, trapped ions, and Rydberg-atom arrays. Once a problem has been expressed in Ising form, it can be implemented directly on a quantum annealer, allowing the optimisation task to be realised as the physical evolution of an interacting quantum system. In this sense, QA offers a fundamentally different approach to tackling computationally hard problems: rather than explicitly searching the configuration space algorithmically, it harnesses the dynamics of a many-body quantum system to explore the underlying energy landscape.

Combinatorial optimisation problems of this type arise across a wide range of applications, including scheduling, logistics, portfolio optimisation, traffic routing, machine learning, and the modelling of complex physical systems. For many of these problem classes, classical heuristics, such as simulated annealing, genetic algorithms, mixed-integer programming, and constraint-programming solvers, are highly effective. However, their performance can degrade rapidly as problem size and complexity increase. QA was originally proposed~\cite{kadowaki1998quantum} as an alternative approach that exploits quantum dynamics to explore such optimisation landscapes.

\begin{figure}[h]
    \centering
    \includegraphics[width=0.43\textwidth]{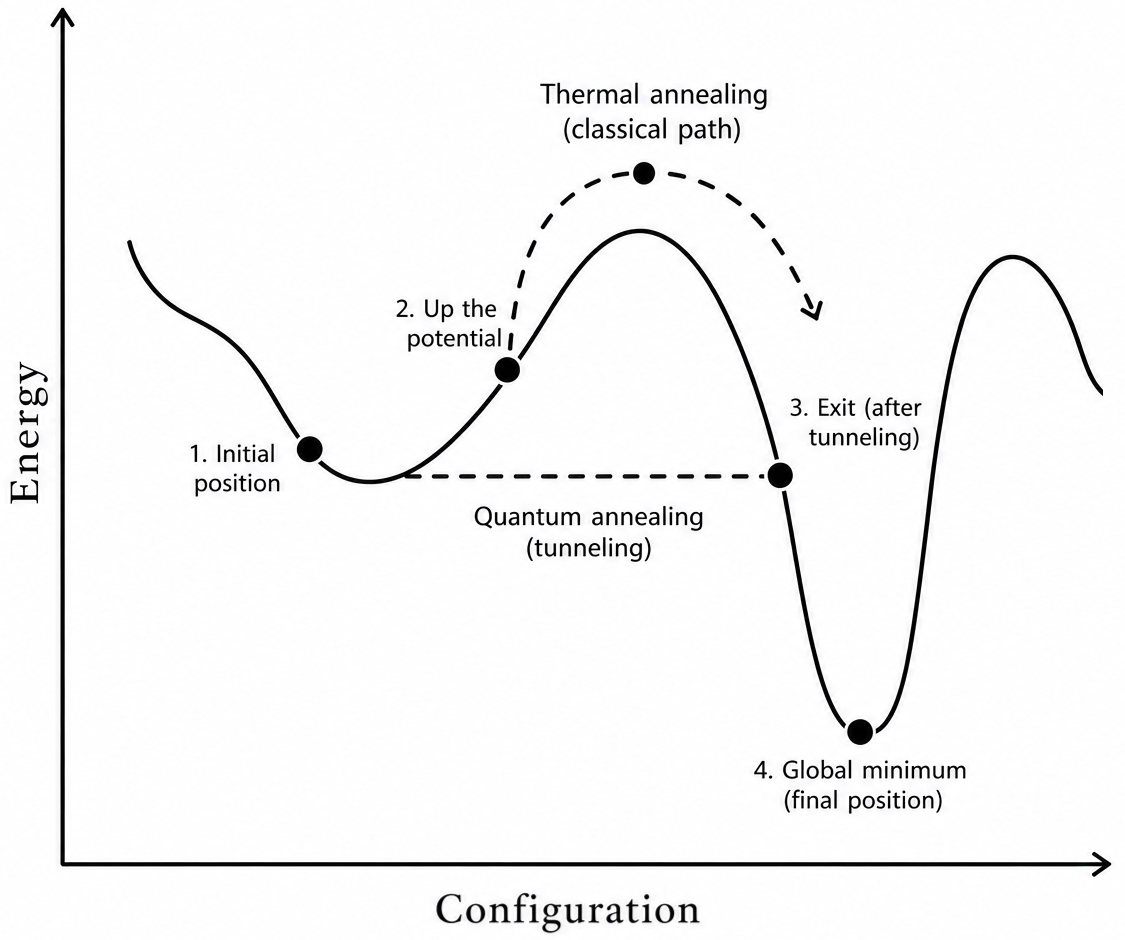}
    \caption{Schematic energy landscape illustrating classical and quantum escape mechanisms. A classical process escapes a local minimum by activation over an energy barrier, while quantum tunnelling can connect configurations through the barrier.}
    \label{fig:energy_landscape_cartoon}
\end{figure}

A distinctive feature of QA, compared with purely classical stochastic search, is quantum tunnelling. In classical algorithms, escaping a local minimum typically requires climbing over an energy barrier through thermal or random fluctuations. A quantum system, by contrast, can transition between configurations even when they are separated by such barriers. 
This mechanism can, in certain situations, alter how the system explores complex energy landscapes, as illustrated schematically in Fig.~\ref{fig:energy_landscape_cartoon}, where classical thermal activation is contrasted with quantum tunnelling across energy barriers. While it does not guarantee improved performance, it highlights a qualitative difference between quantum and classical search dynamics that underlies much of the interest~in~QA.

\subsection{Why pay attention to QA?}

The qualitative differences between classical and quantum dynamics raise a central question: can these differences translate into a practical computational advantage? In particular, can QA solve optimisation problems that remain difficult for the best available classical methods? At present, there is no evidence for a universal quantum advantage in solving real-world optimisation problems, which is consistent with the expectation that many such problems remain hard even for quantum computers.

Nevertheless, quantum annealers exhibit capabilities that make them of both computational and physical interest. Modern devices implement programmable systems containing thousands of qubits, placing them among the largest controllable quantum systems currently available. In the idealised limit of perfectly coherent and sufficiently slow evolution, QA reduces to adiabatic quantum computation, which is computationally equivalent to the circuit model. In practice, however, existing devices operate in open-system regimes where coherent evolution, thermal effects, and noise all play a role. They are therefore best understood as physical systems that generate low-energy configurations through a combination of quantum and classical dynamics.

Within this framework, several studies report that QA and hybrid quantum--classical approaches can perform competitively with leading classical heuristics for certain structured optimisation tasks, particularly when the objective is to obtain good approximate solutions within limited time rather than provably exact answers. In such hybrid approaches, classical algorithms generate candidate solutions or decompose large problems into smaller components, while the annealer explores difficult regions of the search space.

At the same time, quantum annealers are valuable beyond optimisation. Because they realise programmable networks of interacting spins, they provide platforms for studying the dynamics of many-body quantum systems and for probing regimes that are difficult to simulate accurately using classical methods. When an optimisation problem is expressed in Ising form, its cost function becomes the energy of a physical system whose parameters are directly implemented in hardware, and the search for low-cost configurations corresponds to its dynamical evolution.
Understanding both the algorithmic performance and the underlying physical dynamics of these systems is therefore essential for assessing their capabilities. For this reason, QA is studied not only as a computational tool, but also as a controllable setting for exploring many-body quantum dynamics. For reviews from both algorithmic and physical perspectives, see Refs.~\cite{AlbashLidar2018, hauke2020perspectives, rajak2023quantum}.

\subsection{Examples of Ising encodings}
\begin{figure}[h]
\centering
\begin{tikzpicture}[scale=.8,
  line/.style={draw=black, line width=1pt},
  v/.style={circle, draw=black, minimum size=4mm, inner sep=0pt},
  blackv/.style={v, fill=black},
  whitev/.style={v, fill=white},
  centerv/.style={v, fill=gray!15}
]

% -------- Left diagram --------
\begin{scope}
  % coordinates (upright square)
  \coordinate (TL) at (0,2);
  \coordinate (TR) at (2,2);
  \coordinate (BL) at (0,0);
  \coordinate (BR) at (2,0);
  \coordinate (C)  at (1,1);

  % edges
  \draw[line] (TL) -- (BL);
  \draw[line] (BL) -- (BR);
  \draw[line] (C) -- (TL);
  \draw[line] (C) -- (TR);
  \draw[line] (C) -- (BL);
  \draw[line] (C) -- (BR);

  % vertices
  \node[blackv]  at (TL) {};
  \node[blackv]  at (TR) {};
  \node[whitev]  at (BL) {};
  \node[blackv]  at (BR) {};
  \node[whitev] at (C)  {};
\end{scope}

% -------- Right diagram --------
\begin{scope}[xshift=4cm]
  \coordinate (TL) at (0,2);
  \coordinate (TR) at (2,2);
  \coordinate (BL) at (0,0);
  \coordinate (BR) at (2,0);
  \coordinate (C)  at (1,1);

  \draw[line] (TL) -- (BL);
  \draw[line] (BL) -- (BR);
  \draw[line] (C) -- (TL);
  \draw[line] (C) -- (TR);
  \draw[line] (C) -- (BL);
  \draw[line] (C) -- (BR);

  \node[whitev]  at (TL) {};
  \node[blackv]  at (TR) {};
  \node[blackv]  at (BL) {};
  \node[whitev]  at (BR) {};
  \node[whitev] at (C)  {};
\end{scope}

\end{tikzpicture}
 \caption{An example of a network problem in which no two coloured nodes may be linked. The coloured nodes on the left form a maximum independent set.\label{fig:netwwork}}
\end{figure}
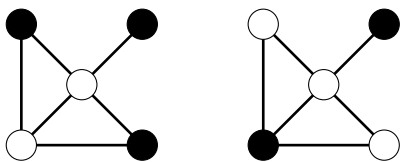

{\bfseries Example 1: a network problem.} As a concrete illustration of how a combinatorial optimisation problem can be encoded in an Ising Hamiltonian, consider the simple network shown in Fig.~\ref{fig:netwwork}. The task is to select the largest possible subset of nodes such that no two selected nodes are directly connected by an edge. This problem is known in graph theory as the \emph{maximum independent set} problem. Two possible choices are shown in the figure below; the configuration on the left corresponds to an optimal solution for this particular network.

The maximum independent set problem is NP-hard; a brute-force approach would require checking all possible subsets of nodes, a procedure whose cost grows exponentially as $2^N$ for a network with $N$ nodes.

To encode this problem in an Ising Hamiltonian, we associate a binary spin variable $s_i=\pm1$ with each node $i$ of the graph. We associate a binary spin variable $s_i=\pm1$ with each node, where $s_i=+1$ denotes that the node is selected. Maximising the number of selected nodes is implemented by a uniform negative bias favouring $s_i=+1$, while the constraint that adjacent nodes cannot both be selected is enforced by adding an energy penalty for each connected pair simultaneously in the $+1$ state. The latter can be achieved by including, for each linked pair $\{i,j\}$, a term proportional to $(1+s_i)(1+s_j)$. This expression vanishes whenever at least one of the two nodes is unselected, but contributes a positive energy penalty when both spins are $+1$. The resulting problem
Hamiltonian can therefore be written as
\begin{equation*}
H_{\text{Ising}}(\{s_i\}) ~=~ -\Lambda \sum_i s_i ~+~\Lambda'\!\!\!\!\!\!\! \sum_{i<j, ~(i,j)\in E} (1{+}s_i)(1{+}s_j),
\end{equation*}
where $E$ is the set of edges of the graph and $\Lambda, \Lambda'$ are positive constants chosen such that the penalty terms dominate over the bias terms. For example, choosing $\Lambda < 4\Lambda'$ ensures that violating the adjacency constraint is always energetically unfavourable. 
To see this, note that flipping a single spin from $-1$ to $+1$ can reduce the bias term by at most $2\Lambda$, whereas if that flip creates a forbidden adjacent pair it increases the penalty by at least $4\Lambda'$ because $(1+s_i)(1+s_j)$ changes from $0$ to $4$ when both spins become $+1$. 
Network problems can be encoded in this manner, and the same idea extends naturally to problems defined on regular lattices. 

This construction illustrates a general principle: combinatorial constraints can be enforced energetically through penalty terms, allowing discrete optimisation problems to be reformulated as ground-state searches of interacting spin systems.
\vspace{8pt}

{\bfseries Example 2: an interface-minimisation problem.} 
As a second example, suppose there are $N^2$ students seated at $N^2$ desks arranged on an $N\times N$ square grid, and exactly half of the students are ill. We would like to assign seats so that as few ill students as possible sit adjacent (horizontally or vertically) to healthy students. This is a simple model of a \emph{segregation} or \emph{interface-minimisation} problem: we wish to minimise the length of the boundary between the two ``types''.
Intuitively, if ill and healthy students are randomly mixed, the boundary between the two groups will be extensive: many adjacent desks will contain one ill and one healthy student. By~contrast, if the ill students cluster together in a compact region of the grid, the number of ill--healthy contacts is reduced. The problem therefore favours large contiguous domains rather than fine-grained mixtures.

We associate a binary variable $s_{i,j}=\pm1$ with each desk at row $i$ and column $j$ of an $N\times N$ square grid, where $s_{i,j}=+1$ denotes an ill student and $s_{i,j}=-1$ a healthy student. The objective is to penalise situations in which ill and healthy students sit next to each other. A convenient choice for the Ising Hamiltonian is
\begin{equation}
H_{\text{Ising}} = H_P + H_C,
\end{equation}
where $H_P$ penalises adjacent desks occupied by students of different type and $H_C$ enforces the constraint that exactly half of the students are ill.
For the penalty term, we sum over nearest-neighbour pairs $\langle u,v\rangle$ on the grid:
\begin{equation}
H_P ~=~ \Lambda \sum_{\langle u,v\rangle}\bigl(1 - s_u s_v\bigr),
\label{eq:HP_seating}
\end{equation}
with $\Lambda>0$. Here $u$ and $v$ label neighbouring desks, and $s_u s_v=+1$ if the two neighbours are of the same type, while $s_u s_v=-1$ if one student is ill and the other healthy. Thus the factor
$(1-s_u s_v)$ vanishes for like neighbours and equals $2$ for unlike neighbours, so $H_P$ counts (up to a constant factor) the number of ill--healthy adjacencies.
To enforce that exactly half the students are ill, note that
\begin{equation}
\sum_{i=1}^N\sum_{j=1}^N s_{i,j}
\end{equation}
equals the number of ill students minus the number of healthy students. The constraint ``half ill, half healthy'' therefore corresponds to $\sum_{i,j} s_{i,j}=0$, which can be imposed energetically by adding
\begin{equation}
H_C
~=~
\Lambda'\left(\sum_{i=1}^N\sum_{j=1}^N s_{i,j}\right)^2,
\label{eq:HC_seating}
\end{equation}
with $\Lambda'>0$. Choosing $\Lambda'$ sufficiently large ensures that any configuration violating the population constraint is energetically disfavoured, so that the ground state minimises the number of unlike adjacencies subject to having exactly $N^2/2$ ill students. 

In the language of statistical mechanics, this construction corresponds to a ferromagnetic Ising model with a fixed total magnetisation. The ground state exhibits phase separation into macroscopic domains of opposite spin, with an energy proportional to the length of the interface between them. This connection highlights how optimisation problems encoded in Ising form can often be interpreted as familiar models of many-body physics.

\vspace{8pt}

{\bfseries Logical vs physical geometry in Ising encodings.}
In the previous example, unlike the local term~$H_P$, which couples only neighbouring desks on the square grid, the constraint term $H_C$ is global. Expanding the square in Eq.~\eqref{eq:HC_seating} produces pairwise couplings between many spins, so that the resulting \emph{logical} Ising Hamiltonian is no longer local on the grid of desks. 
This example highlights an important distinction between three geometries in QA: (i) the structure of the original optimisation problem, (ii) the logical interaction graph defined by the Ising encoding, and (iii) the physical connectivity of the hardware.

In the present example, if only the nearest-neighbour term $H_P$ were present, the logical interaction graph would coincide with the square grid of desks: each spin would interact only with its immediate neighbours. However, the addition of the global constraint $H_C$ changes this situation. Expanding Eq.~\eqref{eq:HC_seating} generates effective couplings between many pairs of spins that are not neighbours on the grid. As a result, the logical interaction graph becomes dense, even though the underlying problem geometry remains two-dimensional.

Annealing hardware, by contrast, implements only local pairwise couplings between qubits that are physically connected on the device. If two spins that are coupled in the logical Hamiltonian are not directly connected in the hardware connectivity graph, the interaction must be realised indirectly. This can be done, for example, by introducing auxiliary qubits or by forming chains of ferromagnetically coupled physical qubits that behave collectively as a single logical variable and mediate the desired interaction. This procedure is known as \emph{minor embedding}: the logical interaction graph of the problem is mapped onto the connectivity graph of the quantum processing unit.
Finding efficient embeddings is itself a nontrivial optimisation problem and can introduce significant overhead in the number of physical qubits required to represent a given logical variable~\cite{Gomez-Tejedor:2025eqs}. Thus, while global constraints can always be expressed using quadratic penalty terms, doing so may transform a geometrically simple optimisation problem into a dense logical interaction graph that is costly to realise on hardware with restricted connectivity. Recent work has explored automated strategies for discovering efficient embeddings, including approaches based on reinforcement learning~\cite{Nembrini:2025fiu}.

This separation between problem structure, logical encoding, and hardware connectivity is a central practical constraint in QA, and plays a key role in determining performance and scalability.

\subsection{Higher-order optimisation problems and quadratic reductions}

The examples discussed so far involve optimisation problems that can be written directly in terms of pairwise interactions between binary variables, leading naturally to Ising Hamiltonians of the form~\eqref{eq:ising}. At first sight this might appear to limit the range of problems that can be addressed, since many optimisation tasks of practical interest involve interactions among three or more variables. However, this restriction is not fundamental. A wide class of higher-order (non-quadratic) optimisation problems can be systematically reduced to equivalent quadratic forms and thus encoded on QA hardware.

The origin of the quadratic restriction lies in the structure of current annealing devices: their programmable Hamiltonians contain only linear terms and pairwise couplings between spins. Interaction terms involving three or more variables therefore cannot be implemented directly. The key idea that overcomes this limitation is to introduce additional binary variables, often referred to as \emph{auxiliary} or \emph{ancilla} variables, that represent products of the original variables. Higher-order interaction terms are then replaced by quadratic interactions involving these auxiliary variables, together with penalty terms that energetically enforce the desired logical relationships. Provided the penalty strengths are chosen sufficiently large, the resulting quadratic Hamiltonian has the same ground states as the original higher-order problem. In this sense the two formulations are \emph{equivalent}: they share the same optimal solutions even though their mathematical representations differ. Comprehensive surveys of such \emph{quadratisation} techniques, and of the associated overhead in auxiliary variables, can be found in Ref.~\cite{dattani2019quadratization}.

To illustrate the principle, consider a simple cubic interaction involving three binary spin variables, proportional to $s_1 s_2 s_3$. To implement this, one may introduce an auxiliary binary variable that represents the product of two spins, for example $s_{12}=s_1 s_2$. The original three-body interaction can then be replaced by a quadratic interaction between this auxiliary variable and the remaining spin, $s_{12}s_3$. In addition, a carefully designed quadratic penalty term is added to enforce the constraint $s_{12}=s_1 s_2$. If the penalty strength is chosen sufficiently large, configurations that violate this constraint become energetically unfavourable, and the lowest-energy states of the quadratic Hamiltonian coincide with those of the original cubic problem. An explicit construction of such penalty terms is given in Appendix~\ref{sec:HigherOrderReduction}.

By applying this construction iteratively, any polynomial optimisation problem over binary variables can in principle be transformed into an equivalent quadratic Ising or QUBO form. While this generally increases the number of variables and couplings in the problem, it greatly enlarges the class of optimisation tasks that can be represented within the Ising framework and hence implemented on QA hardware. A more detailed discussion, including worked examples of these reductions, is presented in Appendix~\ref{sec:HigherOrderReduction}.

The main lesson is that the pairwise form of present-day annealing Hamiltonians should be viewed as a hardware constraint rather than a fundamental limitation of the framework itself. Higher-order cost functions and constraints can often be absorbed into an enlarged quadratic model, albeit at the price of additional variables, denser logical interactions, and greater embedding overhead. This trade-off between expressive power and physical resource cost is a recurring theme in QA, and will reappear throughout this review.

The examples and constructions discussed above illustrate the basic logic of QA. A discrete optimisation problem is first rewritten as the ground-state problem of an Ising or QUBO Hamiltonian; constraints are incorporated through interactions and penalty terms; and the resulting logical model is then implemented, exactly or approximately, on a programmable quantum device. In this way, cost functions become energies, candidate solutions become spin configurations, and optimisation is recast as the controlled dynamics of an interacting many-body system. This reformulation is what gives QA its distinctive character: it is simultaneously a computational strategy and a physical framework.

\medskip
The remainder of this review develops this perspective in stages. We begin in Sec.~\ref{sec:principles} by introducing the theoretical foundations of QA, including adiabatic evolution, spectral gaps, tunnelling, and the role of open-system effects. We then turn to the main hardware platforms and practical considerations that shape present-day implementations, such as connectivity constraints, embedding overhead, control precision, and schedule design. Building on this foundation, we survey algorithmic developments and benchmark questions, with particular emphasis on the distinction between exact optimisation, approximate optimisation, and sampling tasks. We then discuss representative applications ranging from optimisation and machine learning to the use of quantum annealers as programmable platforms for exploring non-equilibrium many-body physics. Throughout, our aim is to present QA as both a computational heuristic and a controllable physical system, and to assess its prospects in that broader setting.

\section{Principles of quantum annealing}
\label{sec:principles}

\subsection{From classical to quantum annealing}

Classical simulated annealing is inspired by the behaviour of a material that is heated and then cooled slowly. At high temperature the system explores many configurations through thermal fluctuations; as the temperature is lowered, it tends to settle into configurations of lower energy. In optimisation, this idea is implemented by introducing a fictitious temperature that allows random moves between candidate solutions and then gradually reducing it so as to favour low-cost configurations.

QA follows the same broad logic, but replaces thermal exploration with quantum dynamics. Instead of relying on temperature to help the system escape local minima, one introduces a quantum \emph{driver} Hamiltonian that mixes classical configurations and can, in favourable situations, enable transitions through energy barriers by tunnelling. The strength of these quantum fluctuations is then gradually reduced, with the aim of steering the system toward low-energy configurations of the target optimisation problem.

The basic mechanism of QA can be summarised as follows: the optimisation problem is encoded in a final Hamiltonian whose ground state represents the desired solution, while the system is initially prepared in the known ground state of a simple non-commuting Hamiltonian. The computation then consists of continuously interpolating between these two Hamiltonians.

To implement this idea, one first promotes the classical Ising energy function~\eqref{eq:ising} to a quantum Hamiltonian acting on qubits by replacing each classical spin variable $s_i=\pm1$ with the operator $\sigma_i^z$. The resulting \emph{problem Hamiltonian} is
\begin{equation}
H_{\text{problem}} = \sum_i h_i \, \sigma_i^z + \sum_{i<j} J_{ij} \, \sigma_i^z \sigma_j^z .
\end{equation}
Because this Hamiltonian contains only $\sigma^z$ operators, it is diagonal in the computational ($\sigma^z$) basis. Its eigenstates therefore correspond directly to classical spin configurations, and their eigenvalues are precisely the associated Ising energies. Finding the ground state of $H_{\text{problem}}$ is thus equivalent to solving the original classical optimisation problem. The difficulty, of course, is that this ground state is not known in advance.

\begin{figure}[t]
\centering
\includegraphics[width=.9\columnwidth]{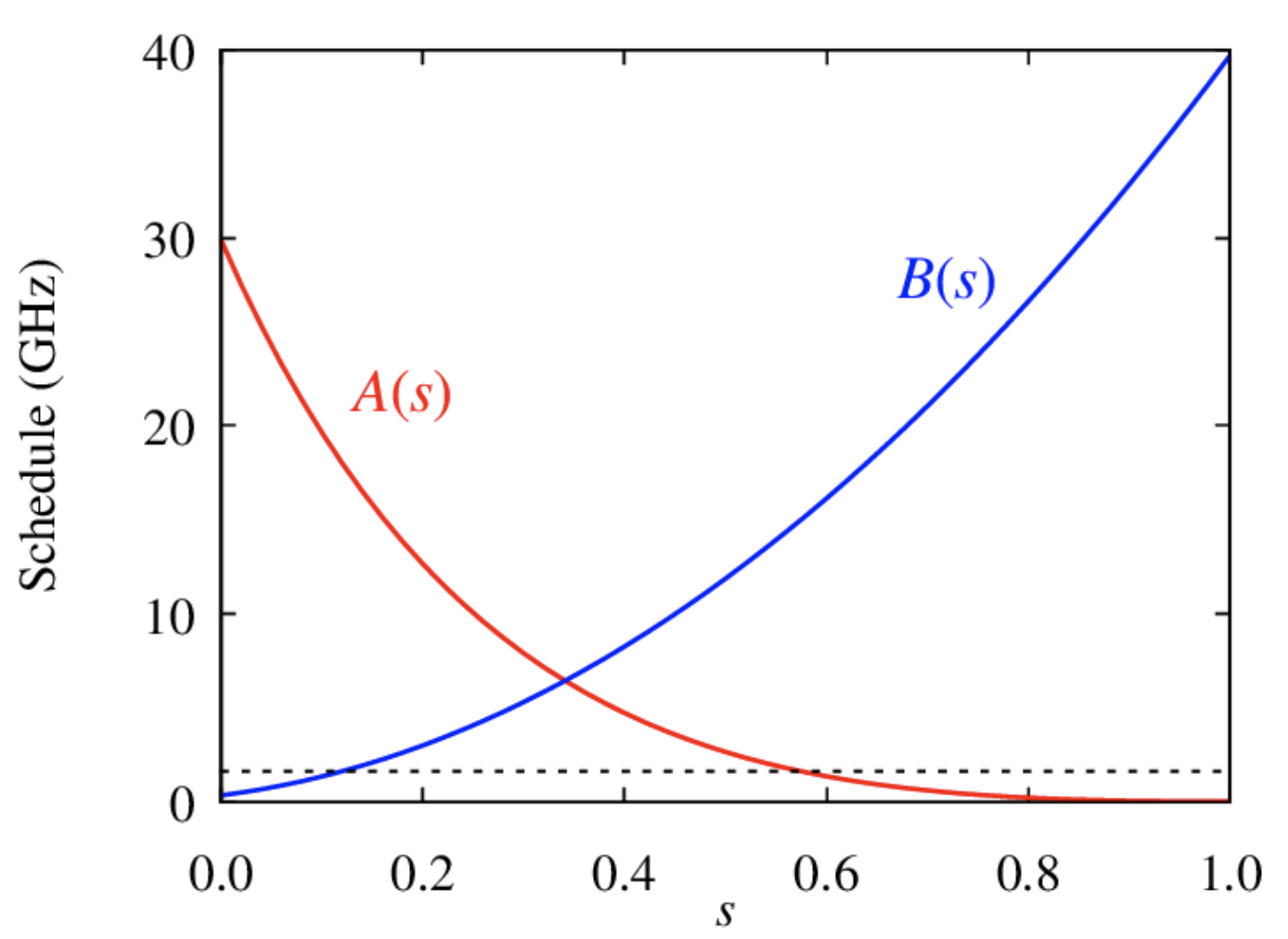}
\caption{Typical annealing schedule functions. The coefficient $A(s)$ controls the strength of the transverse-field driver, while $B(s)$ scales the problem Hamiltonian. In experimental implementations, these coefficients correspond to physical energy scales (often expressed in GHz for superconducting devices). The~evolution interpolates from a regime dominated by coherent superposition over computational-basis states at $s=0$ to one in which the Hamiltonian is diagonal in the computational basis at $s=1$, so that measurements yield classical configurations corresponding to candidate solutions.}
\label{fig:Anneal-schedule-parameters}
\end{figure}

QA addresses this by beginning from the ground state of a much simpler \emph{driver Hamiltonian}, chosen so that it is easy to prepare and does not commute with $H_{\text{problem}}$. A standard choice is a transverse field,
\begin{equation}
H_{\text{driver}} = -\sum_i \Delta_i \, \sigma_i^x ,
\end{equation}
where $\sigma_i^x$ is the Pauli--$x$ operator on qubit $i$. For $\Delta_i>0$, the ground state of $-\Delta_i\sigma_i^x$ is $(\ket{0}+\ket{1})/\sqrt{2}$, and for $n$ qubits the ground state of $H_{\text{driver}}$ is the tensor product of these states, i.e.~an equal-amplitude superposition of all $2^n$ computational-basis configurations. This delocalised state is what is meant, in this context, by quantum fluctuations.
The annealing process is generated by a time-dependent Hamiltonian
\begin{equation}
H(t)=A(t)\,H_{\text{driver}}+B(t)\,H_{\text{problem}},
\label{eq:Ht}
\end{equation}
with schedule functions satisfying
\begin{equation}
A(0)\gg B(0)\approx 0, \qquad A(\tau)\approx 0 \ll B(\tau),
\end{equation}
where $\tau$ is the total annealing time. A typical form of these schedule functions is shown in Fig.~\ref{fig:Anneal-schedule-parameters}. It is convenient to introduce a dimensionless parameter $s=t/\tau\in[0,1]$ that tracks the progress of the anneal.

At $s=0$ the Hamiltonian is dominated by the transverse-field driver, whose ground state is an equal-amplitude superposition of all computational-basis configurations. In this regime the system is delocalised over configuration space due to the non-commuting $\sigma^x$ terms, which coherently mix different $\sigma^z$ configurations. As $s$ increases, the influence of the problem Hamiltonian grows while the transverse field is gradually reduced. By $s=1$ the Hamiltonian is dominated by the problem term and is diagonal in the computational basis, so transitions between configurations are suppressed. Measurements at this stage yield classical bitstrings that correspond to candidate solutions of the optimisation problem.

If this interpolation is performed sufficiently slowly and the system remains close to the instantaneous ground state throughout the evolution, the final measurement yields a ground state, or at least a low-energy state, of $H_{\text{problem}}$. This is the adiabatic picture underlying QA and forms the basis of adiabatic quantum computation. 

A key condition for the adiabatic approximation is that the minimum spectral gap along the evolution remains finite. If the gap becomes very small, the required annealing time increases, and in the limit of a vanishing gap the adiabatic approximation formally breaks down. In practice, however, quantum annealing is often operated outside the strictly adiabatic regime: non-adiabatic transitions, thermal relaxation, and repeated sampling can still lead to useful low-energy configurations. As a result, performance is not determined solely by the minimum gap, but by a more complex interplay between spectral structure, dynamics, and environmental effects. Geometric phases acquired during adiabatic evolution, such as Berry phases, do not affect measurement outcomes in the computational basis and therefore typically play no direct role in optimisation applications of QA.

A useful physical picture is that the driver Hamiltonian initially delocalises the quantum state over many configurations, effectively smoothing the classical energy landscape. As the ratio $A(t)/B(t)$ decreases, the structure of the problem Hamiltonian becomes progressively more important, and the low-energy eigenstates become increasingly concentrated around favourable classical configurations. Because the transverse-field term mixes computational-basis states, probability amplitude can flow between configurations during the anneal. In this sense, the system can move between regions of configuration space separated by barriers in the classical landscape, a mechanism often described as quantum tunnelling.

Present-day quantum annealers operate in a less ideal regime. They are open quantum systems, subject to decoherence, control errors, and thermal noise, and many practically interesting problems exhibit small spectral gaps that make perfectly adiabatic evolution difficult. As a result, real devices typically function neither as perfectly coherent adiabatic computers nor as purely classical optimisers, but in an intermediate regime where coherent dynamics, tunnelling, dissipation, and thermal relaxation all play a role.

For this reason, the output of a practical quantum annealer is often better understood not as ``the ground state'' itself but as a distribution over low-energy configurations produced by noisy, finite-time many-body dynamics. In many devices one observes an effective freeze-out stage beyond which the state changes only weakly, so that the measured bitstrings resemble samples from an approximately thermal distribution of an effective classical Ising model. Whether this is a limitation or a useful feature depends on the task: for exact ground-state optimisation it is an obstacle, whereas for sampling-based applications it can be precisely the desired behaviour. This dual interpretation of QA, as a ground-state optimisation strategy and as controlled many-body dynamics, will recur throughout the rest of this review.

\subsection{Adiabatic quantum computation, spectral gaps, and universality}
\label{sec:AQCuniversality}

In the idealised limit of closed-system dynamics and perfectly slow interpolation, QA reduces to \emph{adiabatic quantum computation} (AQC). In this setting, the computation is implemented by preparing the ground state of a simple initial Hamiltonian and then varying the Hamiltonian slowly enough that the evolving state remains close to the instantaneous ground state throughout the process. Whether this ideal adiabatic picture holds is governed by the quantum adiabatic theorem and, in particular, by the behaviour of the spectral gap along the annealing path.
Let $H(t)$ be a time-dependent Hamiltonian defined for $t\in[0,\tau]$, with instantaneous eigenvalues and eigenvectors
\[
H(t)\ket{\psi_k(t)}=E_k(t)\ket{\psi_k(t)}, \qquad E_0(t)\leq E_1(t)\leq \cdots.
\]
Assuming for the moment that the ground state remains non-degenerate throughout the evolution, one defines the instantaneous gap
\begin{equation}
\Delta(t)=E_1(t)-E_0(t),
\end{equation}
and the minimum gap along the interpolation,
\begin{equation}
\Delta_{\min}=\min_{t\in[0,\tau]}\Delta(t).
\end{equation}

\vspace{4pt}
\noindent{\bfseries Quantum Adiabatic Theorem (informal statement).}
{\itshape
If the Hamiltonian varies smoothly in time, the system is prepared initially in the ground state, and the minimum spectral gap remains nonzero, then sufficiently slow evolution keeps the state close to the instantaneous ground state throughout the interpolation. Quantitatively, the deviation is controlled by a ratio of the form
\[
\frac{\max_{t\in[0,\tau]}\|\dot H(t)\|}{\Delta_{\min}^2}.
\]
}

Rigorous versions of the theorem make the smoothness assumptions and constants precise~\cite{Jansen2007,AlbashLidar2018}. The essential message is that adiabatic evolution becomes difficult when the gap becomes small: narrow avoided crossings act as dynamical bottlenecks, requiring slower interpolation to suppress transitions out of the ground state.

This can be understood by considering how the spectrum evolves during the anneal. In generic finite-dimensional systems depending smoothly on a single parameter, exact level crossings are rare unless protected by symmetry. More commonly, one encounters \emph{avoided crossings}, where two levels approach closely but do not become degenerate. Near such a point the dynamics is often well approximated by an effective two-level system. If the avoided crossing is traversed at a rate $v$, the Landau--Zener formula gives an excitation probability of the form
\begin{equation}
P_{\mathrm{exc}} \sim \exp\!\left(-\frac{\pi \Delta^2}{2v}\right),
\end{equation}
where $\Delta$ is the minimum gap. This makes explicit why small gaps control the runtime of adiabatic algorithms.

The minimum spectral gap is therefore the key quantity linking the physical dynamics of the anneal to its computational complexity. As the number of variables increases, the Hilbert space grows exponentially, and the spectrum must reorganise an exponentially large set of competing configurations during the interpolation. For hard optimisation problems, one generally expects the relevant low-energy states to become increasingly crowded, often leading to very small avoided crossings. Although there is no general theorem that NP-hardness implies exponentially small gaps for every annealing path, the heuristic expectation is that difficult optimisation landscapes are frequently accompanied by severe spectral bottlenecks.

A further subtlety arises when the ground state is degenerate. Many optimisation problems possess multiple optimal solutions, so the final Hamiltonian may have a degenerate ground-state manifold. In such cases the simplest non-degenerate adiabatic theorem does not apply globally. Several possibilities may occur: the ground-state subspace may remain separated from excited states by a finite gap throughout the evolution; degeneracy may appear only at the final time; or one may introduce a small symmetry-breaking perturbation to lift an otherwise exact degeneracy. In each case, the physical conclusion is similar: sufficiently slow evolution can still guide the system toward the low-energy manifold, but the simplest adiabatic picture must be interpreted with care.

In the fully coherent limit, this adiabatic framework defines a model of universal quantum computation. A central theoretical result is that AQC and the standard circuit model are computationally equivalent~\cite{aharonov2008}. Any quantum circuit can be efficiently encoded into an adiabatic evolution, and conversely any adiabatic computation can be efficiently simulated by a sequence of quantum gates, with only polynomial overhead. In one direction, the construction proceeds by introducing a ``clock'' register and building a Hamiltonian whose ground state encodes the \emph{history state} of the circuit,
\begin{equation}
\frac{1}{\sqrt{T+1}}\sum_{t=0}^{T}\lvert\psi_t\rangle\otimes\lvert t\rangle,
\end{equation}
where $\lvert\psi_t\rangle$ is the computational state after $t$ gates. In the other direction, the continuous adiabatic evolution can be approximated by sufficiently fine discrete time steps and therefore simulated within the circuit model.

This equivalence shows that, in the ideal adiabatic limit, QA inherits the full computational power of quantum computation. It does \emph{not}, however, imply that generic NP-hard optimisation problems become efficiently solvable, nor should it be conflated with the capabilities of present-day quantum annealers. The universality result concerns an ideal closed-system model with sufficiently rich Hamiltonian control and sufficiently slow evolution. Existing annealing devices operate in more restricted, noisy, and often open-system regimes, and are therefore best understood as specialised physical optimisers and low-energy samplers rather than as universal adiabatic quantum computers.

The distinction between ideal AQC and practical QA is therefore much like the distinction between fault-tolerant gate-based quantum computation and present-day noisy circuit devices. The ideal limit provides an essential theoretical benchmark, but the practical behaviour of real annealers is shaped by finite temperature, dissipation, control noise, and schedule design. Understanding that intermediate regime is crucial both for benchmarking optimisation performance and for interpreting quantum annealers as controllable many-body dynamical systems.

\subsection{Protocols and dynamical regimes of quantum annealing}

The discussion so far has focused on the idealised limit of closed-system, fully adiabatic evolution. In practice, however, the term \emph{quantum annealing} refers to a broader family of related protocols that share a common Hamiltonian structure but differ in their schedule design, degree of coherence, and physical operating regime~\cite{AlbashLidar2018,hauke2020perspectives}. It is therefore useful to distinguish between the \emph{control protocol} applied to the system and the \emph{dynamical regime} that the resulting evolution actually realises.

Once a logical optimisation problem has been mapped onto a set of physical qubits, the device implements a time-dependent Hamiltonian of the general form
\begin{equation}
\mathcal{H}(t)=A\!\bigl(s(t)\bigr)\sum_i \sigma_i^x
+
B\!\bigl(s(t)\bigr)\Bigl(\sum_i h_i\sigma_i^z+\sum_{i<j}J_{ij}\sigma_i^z\sigma_j^z\Bigr),
\label{eq:physical_QA_H}
\end{equation}
where $s(t)\in[0,1]$ is a dimensionless anneal parameter, $A(s)$ controls the strength of the transverse-field driver, and $B(s)$ sets the scale of the problem Hamiltonian. The functions $A(s)$ and $B(s)$ are device-specific but typically satisfy
\begin{equation}
A(0)\gg B(0), \qquad A(1)\ll B(1),
\end{equation}
so that the evolution begins in a driver-dominated regime and ends in one dominated by the Ising problem Hamiltonian. In some platforms, additional controls allow partial reshaping of the Hamiltonian during the anneal, for example by varying the effective scale of the local fields independently of the couplings,
\begin{equation}
\sum_i h_i \sigma_i^z \ \longrightarrow \ C(t)\sum_i h_i \sigma_i^z.
\label{eq:h_gain}
\end{equation}
Such controls make modern annealers closer to programmable analogue simulators than to implementations of a single fixed algorithm.

The simplest protocol is \emph{forward annealing}, in which $s(t)$ increases monotonically from $0$ to $1$. More generally, an \emph{anneal schedule} is any experimentally allowed choice of the control function $s(t)$ together with possible time dependence in additional parameters such as $C(t)$. Different choices of schedule can substantially modify how the system explores the energy landscape, and in realistic devices the resulting behaviour depends not only on the nominal control protocol but also on the interplay between coherence, non-adiabatic transitions, and dissipation.
Several important protocol families arise in practice.

\paragraph{Closed-system adiabatic annealing.}
In the idealised regime of negligible noise and sufficiently slow evolution, the system remains close to the instantaneous ground state throughout the anneal. This is the regime described by the adiabatic theorem and corresponds to adiabatic quantum computation~\cite{Farhi2001,AlbashLidar2018}. Its performance is controlled primarily by the minimum spectral gap encountered along the annealing path.

\paragraph{Open-system analogue quantum annealing.}
Real quantum annealers generally operate as open quantum systems coupled to an environment. Thermal excitation, relaxation, dephasing, control noise, and calibration errors all affect the dynamics of the device. In this regime the device no longer deterministically prepares the ground state, but instead generates a distribution over low-energy configurations shaped by the competition between coherent tunnelling and dissipative relaxation~\cite{amin2015searching}. An important empirical feature of many devices is a \emph{freeze-out} stage beyond which the state evolves only weakly, so that measured outputs resemble samples from an approximately frozen low-energy distribution.

\paragraph{Diabatic and schedule-engineered annealing.}
Strictly adiabatic evolution is often neither achievable nor optimal. If the interpolation is performed at moderate speed, the system may undergo diabatic transitions near avoided crossings and temporarily populate excited states. Although such transitions are errors from the perspective of ideal AQC, they can sometimes assist optimisation by helping the system bypass bottlenecks associated with very small gaps. This has motivated \emph{schedule-engineered} protocols that deliberately exploit controlled non-adiabatic dynamics~\cite{muthukrishnan2016tunneling}. Examples include pauses inserted at selected points in the anneal, non-monotonic schedules that temporarily re-strengthen the driver, and inhomogeneous driving schemes in which different qubits experience different transverse-field profiles.

\paragraph{Reverse annealing and local quantum search.}
In \emph{reverse annealing}, the system is initialised near $s\simeq1$ in a chosen classical configuration, typically obtained from a classical heuristic or from a previous annealing run. The control parameter is then reduced to an intermediate value to reintroduce quantum fluctuations, and subsequently increased again for readout. Rather than exploring the full configuration space from a uniform superposition, the annealer performs a local quantum search around a candidate solution. This strategy is particularly useful in hybrid workflows, where classical routines generate initial guesses and the annealer is used to refine them through a combination of tunnelling and thermal relaxation. Additional controls such as the $h$-gain in Eq.~\eqref{eq:h_gain} can be used to reshape the effective landscape during this local search~\cite{perdomo2011study}.

\paragraph{Digital and hybrid annealing-inspired methods.}
Annealing-type strategies can also be implemented on gate-based quantum processors by digitising the continuous time evolution of Eq.~\eqref{eq:physical_QA_H}, for example through Trotterised or variational protocols. More broadly, hybrid quantum--classical methods use quantum devices to generate candidate solutions or explore restricted subspaces while classical routines handle decomposition, parameter updates, and post-processing. Closely related are several ``quantum-inspired'' platforms, such as coherent Ising machines and specialised digital annealers, which implement Ising or QUBO optimisation without relying on fully coherent many-body quantum evolution. Although their microscopic physics differs from that of quantum annealers, they are often compared within the same optimisation framework.

These distinctions make clear that QA should not be understood as a single algorithm executed in a single physical limit. Rather, it is a family of optimisation and sampling protocols built around a common programmable Hamiltonian, but realised under different assumptions about control, coherence, and environment. Forward adiabatic annealing, open-system analogue QA, diabatic schedule engineering, reverse annealing, and digital or hybrid variants all inhabit different regions of this design space. Understanding which regime is relevant for a given device or application is essential both for fair benchmarking and for interpreting quantum annealers as controllable many-body dynamical systems.

We therefore turn next to the hardware platforms on which these different forms of annealing are realised.

\section{Hardware and Performance}

\subsection{Quantum annealing hardware}

A central question for QA is how the abstract Ising model introduced above is realised in physical hardware. Different experimental platforms implement this model with markedly different trade-offs in system size, interaction connectivity, coherence time, temperature, and controllability, and these differences play a decisive role in determining both performance and application domains. 
It is therefore useful to compare existing platforms along a small number of key axes: system size, connectivity, coherence, and operating regime. A concise comparison is given in Table~\ref{tab:qa_platforms}, which summarises the main trade-offs across current implementations. These distinctions highlight that different platforms are naturally suited to different applications, ranging from large-scale optimisation to controlled studies of quantum many-body dynamics.

\begin{table*}[t]
\centering
\caption{Comparison of major hardware platforms for QA and related Ising-type dynamics. Values are indicative and evolve with experimental progress.}
\label{tab:qa_platforms}
\begin{tabular}{lccc}
\hline\hline
 & \textbf{~~Superconducting~~} & \textbf{~~Rydberg atoms~~} & \textbf{~~Trapped ions~~} \\
\hline\hline\\[-8pt]
\textbf{Physical qubits} 
& $10^3$--$10^4$ 
& $10^2$--$10^3$ 
& $10$--$10^2$ \\[2pt]

\textbf{Connectivity} 
& Sparse (engineered) 
& Geometric (decaying) 
& Long-range \\[2pt]

\textbf{Coherence} 
& Short 
& Intermediate 
& Long \\[2pt]

\textbf{Temperature} 
& Finite ($\sim$mK) 
& Near-zero 
& Near-zero \\[2pt]

\textbf{Typical regime} 
& Open, thermal QA 
& Coherent dynamics 
& Coherent / digital QA \\[2pt]

\textbf{Primary use} 
& Optimisation 
& Simulation / structured QA 
& ~~Simulation / hybrid QA~~ \\

\hline
\end{tabular}
\end{table*}

\medskip
\paragraph{Superconducting flux qubits.}

Networks of superconducting flux qubits currently provide the largest programmable implementations of Ising-type Hamiltonians~\cite{johnson2011quantum}. The transverse field and pairwise couplings are realised through externally controlled magnetic flux biases, allowing direct programming of the coefficients $h_i$ and $J_{ij}$. Processors developed by D-Wave Systems contain thousands of qubits and are accessible through cloud-based interfaces, making them the only large-scale, programmable quantum annealers currently deployed for optimisation tasks.

These devices operate at millikelvin temperatures and are intrinsically open quantum systems: coherence times are limited, and coupling to electromagnetic environments leads to relaxation and decoherence. As a result, their dynamics are well described by open-system quantum evolution at finite temperature.
Their primary strengths lie in scale and programmability. The main limitations arise from restricted native connectivity, control noise, and finite precision in the programmable couplings. Many optimisation problems require interactions between variables that are not directly connected on the hardware graph. In such cases, an embedding procedure is used in which a single logical variable is represented by a chain of several physical qubits in order to reproduce the desired couplings. This increases the total number of qubits required to implement a given problem and can significantly affect performance and scaling~\cite{venturelli2015quantum}.

Successive generations of annealing processors have therefore focused on increasing connectivity. The Pegasus and Zephyr architectures used in recent D-Wave systems provide substantially higher connectivity than earlier designs, reducing embedding overhead and enabling larger logical problem instances to be implemented~\cite{Boothby2020,King2023}.

\medskip
\paragraph{Rydberg-atom arrays.}

Neutral atoms trapped in optical tweezer arrays provide an alternative platform for programmable spin models, with interactions arising from van der Waals forces between atoms excited to Rydberg states~\cite{Labuhn:2016xba, Bernien:2017ubn}. The Rydberg blockade mechanism generates effective interactions between nearby atoms, enabling the implementation of transverse-field Ising--type Hamiltonians. Current experiments involve hundreds of qubits arranged in flexible two-dimensional geometries.

These systems operate at effectively negligible thermal occupation and exhibit largely coherent dynamics over experimentally relevant timescales. Decoherence is dominated by finite Rydberg-state lifetimes and technical laser noise rather than by thermal relaxation.
Rydberg platforms offer high coherence and geometric flexibility, making them particularly well suited to problems with spatial structure and to the study of coherent many-body dynamics. However, the interaction strength depends on the physical distance between atoms and typically decays with separation, making the implementation of fully arbitrary pairwise couplings more challenging than in superconducting devices.

Most neutral-atom processors are developed both in academic settings and by emerging companies such as QuEra Computing and Pasqal. Although these platforms are primarily designed as quantum simulators or gate-based quantum processors, their ability to realise programmable Ising Hamiltonians makes them a natural testbed for annealing-type dynamics and optimisation heuristics.

\medskip
\paragraph{Trapped ions.}

Trapped-ion systems realise effective spin models through laser-induced couplings mediated by collective vibrational modes~\cite{kim2010quantum, Islam:2011btc}. These interactions naturally generate long-range transverse-field Ising Hamiltonians, often with near all-to-all connectivity for modest system sizes. Current implementations typically involve tens of qubits, with ongoing efforts, including those of including Quantinuum and IonQ, toward larger-scale systems.

These platforms operate in a highly coherent regime, with long coherence times and precise control over interactions. Their strengths lie in controllability and flexibility: interaction graphs can be engineered with high fidelity, and both analogue and digital implementations of annealing dynamics are possible.

At present, however, system size remains limited compared with superconducting annealers. As a result, trapped-ion platforms occupy a complementary regime: smaller but highly controllable systems that are particularly valuable for studying annealing dynamics, testing algorithmic ideas, and implementing digital or hybrid annealing-inspired protocols.

\medskip

These platforms realise different points in a multidimensional design space. Superconducting devices prioritise scale and programmability but operate in a strongly open, finite-temperature regime. Rydberg arrays and trapped ions offer higher coherence and greater control, enabling detailed studies of many-body dynamics, but at more modest system sizes. No existing architecture simultaneously optimises scale, connectivity, coherence, and control accuracy, and different applications of QA naturally favour different regions of this design space.

From a user's perspective, however, the high-level workflow is largely platform-independent: encode the problem as an Ising or QUBO Hamiltonian, implement a time-dependent interpolation between driver and problem terms, and sample low-energy configurations through measurement. What distinguishes different implementations is not the mathematical formulation of the optimisation problem, but the physical regime in which the energy landscape is explored, whether predominantly coherent, strongly dissipative, schedule-engineered, or hybrid.
Several academic and industrial efforts continue to develop these complementary platforms, reflecting the broader view of QA as both a computational paradigm and a framework for programmable many-body dynamics.

\medskip

Beyond these architectural considerations, practical use of quantum annealers introduces additional constraints that are often underemphasised in theoretical discussions. On real devices, the programmable coefficients $h_\ell$ and $J_{\ell m}$ must lie within hardware-dependent ranges, and software stacks commonly provide an ``auto-scale'' option that rescales user-specified coefficients to fit these limits. While convenient, autoscaling can unintentionally weaken intended penalty gaps or chain couplings. In practice it is often preferable to choose penalty strengths and embeddings so that the instance fits within native coefficient ranges without global rescaling, and to verify this explicitly when benchmarking or comparing performance across devices.

\subsection{Performance, optimisation, and sampling}
\label{sec:benchmarking}

QA is best understood as a stochastic dynamical process: for a given problem instance, device, and anneal schedule, it produces samples biased toward low-energy configurations rather than a deterministic certificate of optimality. As discussed in the previous section, real devices operate in regimes shaped by coherent tunnelling, non-adiabatic transitions, and dissipative processes. Assessing performance therefore depends both on the computational objective and on the statistical properties of the returned solutions.

A useful organising principle is to distinguish three distinct but closely related performance objectives: exact optimisation, approximate optimisation, and sampling. These correspond to different ways of interpreting the output distribution produced by the annealer.

\medskip
\paragraph{Exact optimisation.}
 %\item \emph{Exact optimisation.} 
 The objective is to obtain a true ground state (or to determine whether the energy lies below a specified threshold) with high probability. The standard performance metric in this regime is the \emph{time-to-solution} (TTS), defined as the expected time required to observe at least one optimal configuration with a chosen target success probability $p$~\cite{ronnow2014defining}. If a single run succeeds with probability $p_s$, the number of repetitions required scales as $\log(1-p)/\log(1-p_s)$, so that TTS depends both on the runtime per anneal and on the success probability per run. Scaling analyses then examine how TTS grows with problem size, typically defined in terms of the number of logical variables. This objective is most closely aligned with the ideal adiabatic picture, where success is governed by the minimum spectral gap and the suppression of diabatic excitations.

\medskip
 \paragraph{Approximate optimisation.}
 %\item \emph{} 
 In many applications it is sufficient to find solutions whose energy lies close to optimal rather than exactly at the minimum. Performance is then quantified by metrics such as the \emph{time-to-$\epsilon$}, defined as the expected time required to obtain a configuration satisfying $E \le E^\star + \epsilon$, where $E^\star$ is the optimal (or best-known) energy. Alternatively, one may report the best, median, or percentile energy reached within a fixed time budget. In combinatorial optimisation it is also common to use the \emph{approximation ratio}, which compares the quality of the returned solution to the optimum. These metrics are often more relevant in practice, since the final steps required to reach exact optimality can demand disproportionately large computational effort.

\medskip
 \paragraph{Sampling.}
 %\item \emph{Sampling.} 
 In some applications the goal is not to minimise energy as aggressively as possible, but to generate configurations according to a useful low-energy distribution. Performance is then evaluated in terms of statistical properties of the samples, such as the frequency of low-energy states, the diversity of solutions, or the effective temperature and correlation structure of the resulting distribution. Sampling objectives arise naturally in probabilistic inference, energy-based machine learning, and in the use of annealers as programmable many-body systems.
%\end{itemize}

Because the success probability can vary widely across instances, benchmarking studies typically consider ensembles of problems rather than individual cases. Across such ensembles, runtime distributions are often broad and may exhibit heavy-tailed behaviour: while many instances are solved relatively quickly, a small fraction of particularly difficult instances can dominate average performance. For this reason, metrics such as TTS are defined statistically and estimated by averaging over many repeated runs and many problem instances~\cite{ronnow2014defining,boixo2014evidence}.

A key practical distinction concerns the difference between \emph{logical} and \emph{physical} problem size. The logical size refers to the number of variables in the original optimisation problem, whereas the physical size corresponds to the number of qubits required after embedding onto the hardware connectivity graph. Because embedding overheads can scale nontrivially with problem structure, meaningful scaling analyses must specify clearly which notion of size is being used.

Reliable performance assessment also requires careful comparison with classical baselines. State-of-the-art classical heuristics, including simulated annealing, parallel tempering, and advanced local-search or branch-and-bound methods, are highly optimised and problem-specific. Fair comparisons must therefore be performed on equivalent problem representations and should account for all relevant computational costs, including embedding, classical pre- and post-processing, and hardware programming time, rather than considering only the raw anneal duration.

A substantial literature has examined whether quantum annealers exhibit a computational advantage over classical optimisation algorithms~\cite{ronnow2014defining,katzgraber2014glassy}. These studies emphasise that claims of ``quantum speedup'' are inherently subtle and depend sensitively on the choice of benchmark instances, the classical algorithms used for comparison, and the performance metric adopted. Distinguishing genuine quantum advantages from artefacts of instance selection or implementation details remains an active area of research.
From this perspective, quantum annealers are best viewed not as deterministic optimisation oracles but as programmable many-body systems that generate structured distributions over low-energy states. Depending on the application, these distributions may be exploited either for optimisation or for sampling. The next section explores how these capabilities are used in practice across a range of applications.

\section{Applications}
\label{sec:applications}

QA occupies an unusual position among quantum technologies. On the one hand, it is motivated by computational optimisation: programmable Ising Hamiltonians provide a common language in which many discrete problems can be expressed, and annealing dynamics offers a physical heuristic for finding low-energy configurations. On the other hand, quantum annealers are also large, controllable many-body systems, which makes them valuable experimental platforms in their own right. As a result, the most compelling applications of QA are not always those in which it already outperforms the best classical solvers end-to-end, but rather those that exploit its distinctive physical and algorithmic characteristics.

A useful unifying perspective is that quantum annealers can be employed in three conceptually distinct roles: 
(i) as heuristic \emph{optimisers} that seek low-energy configurations, 
(ii) as stochastic \emph{samplers} that generate structured distributions over configurations, and 
(iii) as programmable \emph{quantum simulators} of many-body dynamics. 
These roles are not mutually exclusive: a given application may combine elements of all three. Nevertheless, distinguishing them clarifies both the capabilities of QA and the appropriate criteria for evaluating performance in different domains.

From this perspective, the application landscape of QA is broader than the narrow question of whether present-day devices exhibit a universal optimisation speedup. In some settings the main interest lies in identifying signatures of genuinely quantum dynamics; in others it lies in using annealers to access many-body regimes that are difficult to simulate classically; and in still others the device serves as a specialised subroutine within a larger hybrid workflow. The most useful notion of ``application'' therefore depends on whether one is primarily interested in optimisation quality, sampling behaviour, or physical dynamics.

It would be impossible to survey every application area in detail. Instead, we organise this section around a small number of representative themes: 
\begin{itemize}
\setlength\itemsep{2pt}
\item benchmarking-driven studies of quantum dynamics and optimisation performance,
\item QA as a laboratory for non-equilibrium many-body physics,
\item sampling-driven applications in machine learning and inference,
\item mappings of mathematically structured problems to Ising/QUBO form,
and
\item a brief overview of industrial optimisation use cases.
\end{itemize}

\subsection{Benchmarking, quantum signatures, and optimisation performance}

Benchmarking-driven studies of quantum dynamics and optimisation performance aim to identify instances of ``quantum advantage'' or ``beyond-classical behaviour'' in~QA. In the strictest sense, one would like to demonstrate a clear runtime advantage over the best available classical algorithms for a well-defined task. In practice, however, studies of QA often pursue a broader set of objectives. These include identifying signatures of genuinely quantum dynamics, using annealers as programmable many-body systems to explore regimes that are difficult to simulate classically, and developing heuristic optimisation or sampling methods that perform well on structured instance families.

These perspectives are complementary. A study can reveal compelling quantum dynamical effects without implying a practical optimisation speedup, and conversely a useful optimisation heuristic need not provide the cleanest setting in which to isolate quantum-mechanical signatures. We discuss these aspects in turn.

\medskip
\paragraph{Quantum signatures in dynamics: crafted benchmarks and tunnelling-dominated instances.}
A recurring theme in the QA benchmarking literature has been the identification of signatures of genuinely quantum dynamics in annealing hardware. In particular, many studies examine whether the behaviour of the device can be explained by classical thermal processes alone, or whether it reflects mechanisms such as coherent tunnelling or non-adiabatic transitions that have no direct classical analogue~\cite{boixo2014evidence, ronnow2014defining}. 

Carefully constructed benchmark instances, often featuring tall and narrow energy barriers separating local minima, are frequently used to probe these effects. In such settings, multi-qubit tunnelling has been argued to provide a mechanism for traversing energy barriers more efficiently than purely thermal activation~\cite{denchev2016computational}. 
At the same time, alternative classical models have been proposed that reproduce some features of the observed behaviour, highlighting the subtlety of interpreting experimental signatures of quantum dynamics~\cite{Shin:2014irr, smolin2014classical}. The goal of these studies is therefore not only to demonstrate potential computational advantages, but also to clarify the physical mechanisms by which annealing devices explore complex energy landscapes and to identify regimes in which quantum effects play a significant role. These benchmarking efforts form an important bridge between the physics of the devices and their potential computational applications.

\medskip
\paragraph{Beyond-classical simulation: annealers as quantum simulators of disordered spin dynamics.}
Optimisation is not the only ``task'' one can benchmark. Because annealers implement large, programmable quantum spin systems, they can be used to probe coherent or partially coherent dynamics in regimes where classical simulation becomes extremely costly. A recent line of work benchmarks superconducting annealers as quantum simulators of disordered quantum spin models, comparing measured observables to predictions of unitary Schr\"odinger evolution on small instances and then exploring larger regimes that are classically difficult to simulate at comparable accuracy~\cite{King:2022okf,doi:10.1126/science.ado6285}. The significance of such results is not that they solve NP-hard optimisation problems, but that they probe many-body quantum dynamics at scales that are otherwise hard to access.

\medskip
\paragraph{Algorithmic advantage in a restricted regime: approximate optimisation and ``time-to-$\epsilon$'' metrics.}
Some of the clearest optimisation-oriented claims focus explicitly on the \emph{approximate} regime: rather than asking for guaranteed ground states, one asks for solutions within a fixed target gap of the optimum. This is a natural regime for QA because (i) real devices are stochastic samplers, and (ii) many practical applications value high-quality solutions under time constraints more than provable optimality. In such studies the main question is whether QA (or QA plus error suppression and post-processing) achieves favourable scaling for a structured family of instances, compared with a strong classical baseline, for small but nonzero approximation error. Ref.~\cite{Bauza:2024cxw} is an example of this style of claim.

\medskip

These studies suggest that the most convincing near-term claims for QA are likely to be domain- and metric-specific, rather than broad end-to-end speedups across generic optimisation problems.

Having discussed these benchmarking-driven perspectives, we now turn to the broader application domains outlined above. The examples below illustrate in particular the roles of annealers as programmable many-body quantum simulators and as heuristic engines for optimisation and sampling tasks.

\subsection{Quantum annealers as laboratories for non-equilibrium many-body physics}
\label{subsec:applications_physics}

The annealing protocol implements a controlled sweep of a many-body system under a time-dependent Hamiltonian. Even when the device is used ``as an optimiser'', the underlying evolution is a non-equilibrium process involving avoided crossings, freeze-out, and (on some platforms) partially coherent dynamics. This makes annealers natural laboratories for studying dynamical phenomena that are central in condensed-matter and statistical physics.
\medskip

\paragraph{Quantum critical dynamics and glassy systems.}
When an anneal crosses (or passes near) a quantum phase transition, the closing of the gap in the thermodynamic limit implies that adiabaticity becomes progressively harder with system size. The resulting dynamics connects QA directly to the physics of critical slowing down and defect production.  Superconducting platforms have used large programmable spin-glass instances to study coherent critical dynamics and scaling behaviour in regimes far larger than those accessible to exact classical simulation~\cite{King:2022okf}.  From the QA viewpoint, these experiments help clarify which parts of the anneal are dominated by coherent evolution, which by open-system relaxation, and how the competition between these processes depends on schedule and problem structure.

Complementary evidence for coherent quantum critical dynamics has been obtained on trapped-ion quantum simulators implementing programmable long-range transverse-field Ising models~\cite{Islam:2011btc,Zhang:2017kde,Bohnet:2016qql}. In these systems, adiabatic sweeps across quantum phase transitions have enabled direct observation of defect production consistent with Kibble–Zurek scaling (a characteristic signature of non-equilibrium critical dynamics) as well as dynamical phase transitions in chains of tens of ions. Because trapped-ion platforms operate in a regime of high coherence and effectively negligible thermal occupation, they provide a particularly clean realisation of closed-system annealing dynamics, offering an important contrast to the open-system behaviour characteristic of superconducting flux-qubit annealers. 

Similar phenomena have also been explored in programmable Rydberg-atom arrays, which realise transverse-field Ising Hamiltonians with tunable interactions via Rydberg blockade. Large-scale coherent dynamics have been observed in systems of up to fifty atoms and beyond, including direct studies of many-body evolution in driven Ising models~\cite{Bernien:2017ubn, Lienhard:2017xnx, Guardado-Sanchez:2017ali}. These experiments probe the space- and time-dependent growth of correlations following quenches and controlled parameter sweeps, and reveal the formation of ordered and antiferromagnetic domains in one- and two-dimensional geometries.
\medskip

\paragraph{First-order transitions, metastability, and collective tunnelling.}
First-order transitions and metastable states provide another natural arena in which tunnelling can play a prominent role. In QA language, these correspond to situations where the low-energy landscape contains distinct valleys separated by barriers, and the anneal dynamics must transfer probability mass between them. Such settings connect naturally to the language of nucleation and decay of metastable phases.

A particularly striking application of this idea is to emulate tunnelling phenomena familiar from quantum field theory by encoding quasi-continuous degrees of freedom into an Ising Hamiltonian. Using domain-wall encodings of scalar fields~\cite{Chancellor:2019uyu,Abel:2020ebj}, programmable annealers can represent effective potentials with multiple minima and controlled barrier structures. In this framework, the system is prepared in a metastable ``false vacuum'' configuration and then evolved under the annealing dynamics. Experiments on a D-Wave processor demonstrated that the resulting statistics of final configurations reveal tunnelling transitions into the lower-energy ``true vacuum'' state~\cite{Abel:2020qzm}.

\begin{figure*}[!ht]
    \centering
    \vspace{0.05\textwidth}
    \includegraphics[width=1\textwidth]{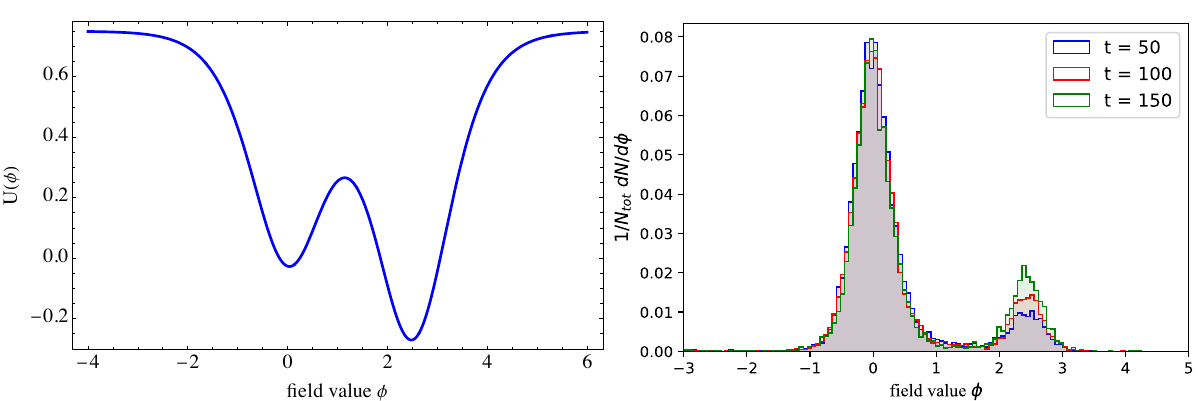}
  \caption{Left: Effective one-dimensional potential $U(\phi)$ realised through a domain-wall encoding, exhibiting a metastable false vacuum near $\phi\simeq 0$ and a lower-energy true vacuum near $\phi\simeq 2.5$. 
Right: Histogram of the measured field values obtained from $N_{\rm tot}$ independent annealing runs. The plotted quantity $(1/N_{\rm tot})\,\mathrm{d}N/\mathrm{d}\phi$ represents the fraction of runs whose final spin configuration corresponds to a field value within a given bin after reverse annealing for a tunnelling time $t$. The emergence and growth of a secondary peak near $\phi\simeq 2.5$ indicates tunnelling transitions from the metastable false vacuum to the true vacuum.}
    \label{fig:vac_dec_pot_and_hist}
\end{figure*}

Operationally, the experiment prepares the encoded field in the metastable minimum and performs repeated annealing runs for different evolution times. Each run produces a spin configuration corresponding to a particular discretised field value, and by collecting many such runs one obtains a histogram of final field configurations, as illustrated in Fig.~\ref{fig:vac_dec_pot_and_hist}. The appearance and growth of a second peak corresponding to the true vacuum directly signals the tunnelling process. The measured tunnelling probabilities were found to agree well with semiclassical expectations based on WKB estimates of instanton-induced decay rates~\cite{Coleman:1977py,Callan:1977pt}. These results demonstrate how annealing hardware can function as a laboratory for observing non-perturbative quantum phenomena that are otherwise difficult to probe experimentally.

Related ideas have recently been explored in programmable spin-chain systems engineered on annealing hardware~\cite{Vodeb:2024tvo}. By embedding hundreds of qubits in carefully designed chain topologies and tuning the couplings to create an effective metastable potential landscape, experiments have observed nucleation-like dynamics associated with the decay of metastable states and the formation of bubble-like structures. Such studies illustrate how large-scale programmable spin systems can emulate collective tunnelling processes in regimes that interpolate between coherent quantum dynamics and thermally assisted relaxation.

\medskip

\paragraph{Other transitions and universality classes.}
Because the programmable Hamiltonians can realise different effective geometries and disorder patterns, annealers have also been explored as platforms for studying dynamical signatures of other universality classes (including experiments that have reported signatures consistent with the Berezinskii--Kosterlitz--Thouless transition in suitably engineered spin systems~\cite{King:2018rbd}) and for testing scaling hypotheses in driven disordered systems.  Such studies highlight the broader potential of annealing hardware as a controllable platform for exploring dynamical phenomena in interacting many-body systems. Whether a given phenomenon is best interpreted as ``an application'' or ``a physics experiment'' is partly semantic; in both cases, the device is used as a controllable many-body dynamical system rather than solely as an optimisation oracle.

\subsection{Machine learning and inference}
\label{subsec:applications_ml}

Many machine-learning tasks can be formulated in terms of energy functions over binary variables, where lower energy corresponds to higher probability under a model distribution. From this perspective, quantum annealers are not only optimisers but also \emph{physical samplers} that return configurations drawn from a hardware-dependent low-energy distribution. This viewpoint connects QA naturally to \emph{energy-based models}, in which learning and inference depend on sampling rather than exact optimisation.

This distinction is essential. In contrast to the optimisation-centric view often adopted in early discussions of QA, many modern applications in machine learning depend not on finding a single ground state, but on generating \emph{representative samples} from a structured distribution. As discussed in Sec.~\ref{sec:benchmarking}, this shifts the relevant performance criteria away from time-to-solution and toward sampling-oriented quantities such as distributional bias, diversity, and effective temperature. In this regime, the central question is not whether QA finds the global optimum, but whether it produces samples that are \emph{useful} for downstream tasks.

A central open problem is whether the distributions generated by quantum annealers differ in a meaningful and exploitable way from those produced by classical sampling algorithms. Because QA dynamics combine quantum tunnelling, non-adiabatic transitions, and thermal relaxation, the resulting samples are not guaranteed to follow a simple Boltzmann distribution. In this sense, the key issue is not whether quantum annealers reproduce an ideal equilibrium sampler, but whether their intrinsically non-equilibrium output distributions constitute a useful computational resource for learning and inference. Understanding this \emph{sampling bias}, and determining when it is beneficial rather than detrimental, remains an active area of research.

\medskip

\paragraph{Boltzmann machines and energy-based models.}
Restricted Boltzmann machines (RBMs) and related energy-based models define probability distributions of the form $p(x)\propto e^{-E(x)}$, where $E(x)$ is typically a quadratic function of binary variables. Training such models requires repeated sampling from both conditional and model distributions in order to estimate gradients of the log-likelihood. Classically, this is often achieved using Markov-chain Monte Carlo methods, which can suffer from slow mixing in rugged energy landscapes.

Quantum annealers offer an alternative sampling mechanism that may explore such landscapes differently, potentially facilitating transitions between modes through a combination of tunnelling and thermal effects. A number of proof-of-principle studies have mapped RBM training and related sampling tasks onto annealers, using the device either to generate candidate samples or to accelerate parts of the training loop~\cite{Adachi:2015fdr,Amin:2016grb,Benedetti:2017ehk,Wilson:2019mqw,xu2021adaptive}. 

The current evidence suggests a nuanced picture. In certain regimes, QA-assisted sampling can be competitive with classical methods, particularly when integrated into hybrid workflows. In others, classical algorithms remain more efficient and robust. The main takeaway is therefore not the existence of a universal speedup, but that QA provides a \emph{distinct sampling primitive} whose statistical properties can differ from those of standard classical samplers. Characterising and exploiting these differences may ultimately be more important than asking whether QA simply reproduces or accelerates conventional classical sampling routines.

\medskip

\paragraph{Classification and weak-learner selection as QUBO.}
Many supervised learning tasks can be reduced to discrete optimisation problems, including feature selection, ensemble construction, and binary classification with sparsity constraints. In such formulations, binary decision variables encode which features or weak learners are selected, and the objective function balances training error against model complexity through regularisation terms. This structure leads naturally to quadratic unconstrained binary optimisation (QUBO) formulations and therefore to implementations on QA hardware.

In this setting, QA functions as a heuristic optimiser within a broader pipeline. Its role is to explore a combinatorial space of candidate models and identify high-quality configurations that can then be evaluated or refined classically. An illustrative example arises in high-energy physics classification problems, where annealing-based optimisation has been used to select and combine weak classifiers constructed from physically meaningful kinematic variables. In small-scale studies, the resulting classifiers achieved performance comparable to classical machine-learning baselines~\cite{Mott:2017xdb}.

More generally, the effectiveness of QA in such tasks depends less on asymptotic scaling than on its ability to navigate structured, high-dimensional search spaces under tight time constraints. This again aligns with the broader view of QA as a heuristic and hybrid tool rather than a standalone replacement for classical optimisation algorithms. More broadly, the long-term impact of QA in machine learning may lie less in replacing established classical methods than in enlarging the space of available sampling and optimisation primitives for complex probabilistic models.

\subsection{Mathematical and algorithmic problems in Ising form}
\label{subsec:applications_math}

A distinctive feature of QA is that it provides a uniform physical realisation of QUBO/Ising instances. This makes it natural to apply annealing-based methods to mathematically structured problems whose core difficulty lies in discrete search under constraints. In such settings, the main task is to reformulate the original problem as the minimisation of an energy function whose ground state encodes a valid solution. From this perspective, QA acts as a general-purpose \emph{physical backend} for discrete optimisation, once a suitable encoding has been identified.
This viewpoint shifts the focus away from individual application domains and toward the broader question of \emph{representability}: which classes of problems can be mapped efficiently to QUBO form, and with what overhead in auxiliary variables, precision, and connectivity. The examples below illustrate this paradigm.

\medskip 

\paragraph{Integer constraints and Diophantine-style searches.}
Many problems involving integer constraints can be mapped to optimisation by introducing penalty terms that enforce the desired conditions. A simple and widely used strategy is to encode an equation $f(x)=0$ by minimising the squared residual $f(x)^2$. Consider, for instance, the search for integer solutions of
\begin{equation}
x^5 + y^5 = z^5 + t^5\, ,
\end{equation}
which asks whether there exist integers that can be written as the sum of two fifth powers in more than one way. Although this problem remains open in mathematics, it can be embedded into an optimisation task through the Hamiltonian
\begin{equation}
H = (x^5 + y^5 - z^5 - t^5)^2 - \lambda(\delta_{x,z}+\delta_{x,t}),
\end{equation}
where $\lambda>0$ penalises trivial solutions.

The practical challenge lies in translating such polynomial expressions into QUBO form. Direct encodings typically generate higher-than-quadratic interactions, which must be reduced to quadratic couplings through the introduction of auxiliary variables (see Appendix~\ref{sec:HigherOrderReduction}). Despite the resulting overhead, this approach enables exploration of extremely large discrete search spaces. For example, Ref.~\cite{Abel:2022wnt} used such encodings to investigate generalisations of Ramanujan's Taxicab numbers and anomaly-cancellation constraints in particle-physics models, scanning configuration spaces of size up to $\mathcal{O}(10^{26})$.

\medskip 

\paragraph{Factoring as an optimisation problem.}
Integer factorisation can likewise be expressed as a quadratic optimisation task. Given a composite number $N$, one may search for factors $p$ and $q$ by minimising
\begin{equation}
H = (N - pq)^2 ,
\end{equation}
over binary encodings of $p$ and $q$, supplemented by constraints that eliminate trivial solutions.

This formulation provides a direct route to implementing factoring on annealing hardware. Using progressively improved encodings and embedding strategies, experiments on D-Wave devices have demonstrated factorisation of integers such as
\begin{equation}
376289 = 571 \times 659 , \qquad
8219999 = 32749 \times 251 ,
\end{equation}
representing some of the largest instances treated on quantum hardware using optimisation-based approaches~\cite{Jiang:2018gut,Ding:2023qty}. These demonstrations should be viewed as proofs of principle rather than competitive algorithms: the asymptotic complexity of optimisation-based factoring is not known, and hardware constraints impose substantial overheads.

It is useful to place these results in context. Implementations of Shor's algorithm on gate-based quantum processors~\cite{Vandersypen_2001, Monz:2015xnh} (the only known polynomial-time quantum algorithm for factoring) have so far been limited to very small integers. Annealing-based approaches therefore explore a complementary regime, in which larger instances can be addressed heuristically, albeit without provable complexity advantages. Importantly, current QA-based methods do not threaten cryptographic systems such as RSA; rather, they provide a framework for studying how optimisation-based quantum hardware processes structured arithmetic problems.

\medskip 

\paragraph{Differential equations via variational encodings.}
Although QA is naturally suited to discrete optimisation, continuous problems such as differential equations can be addressed through discretisation or variational encodings. In such approaches, the solution is expanded in a finite basis,
\begin{equation}
u(x) \approx \sum_k c_k \phi_k(x),
\end{equation}
and the coefficients $c_k$ are determined by minimising a residual norm derived from the governing equation. After discretisation, this optimisation can be mapped to a QUBO problem.

Recent work has proposed general frameworks for solving linear partial differential equations using this strategy~\cite{Criado:2022deo}. In these schemes, the annealer serves as a subroutine for minimising a discretised loss function, while classical post-processing refines the solution and compensates for hardware limitations. From this perspective, QA acts as a specialised optimisation primitive embedded within a broader hybrid numerical pipeline.

\medskip

These examples illustrate a central point: the strength of QA in this context does not lie in any single application domain, but in its ability to provide a common physical framework for a wide class of structured discrete search problems once they have been reformulated in Ising or QUBO form. The key challenges are therefore not only algorithmic but representational, involving trade-offs between encoding efficiency, precision, and hardware constraints. Understanding and optimising these mappings is likely to be as important as improving the underlying annealing dynamics.

\subsection{Industrial optimisation}
\label{subsec:applications_industry}

Most industrial interest in QA is driven by combinatorial optimisation problems such as scheduling, routing, allocation, and matching, in which discrete decisions must be made under constraints~\cite{neukart2017traffic,yarkoni2022quantum}. In these settings, the clean theoretical question ``does QA asymptotically outperform classical solvers?'' is often less relevant than a more practical one: can QA provide \emph{high-quality solutions quickly}, or improve the performance of an end-to-end workflow when used as one component among others?
From this perspective, QA is best viewed not as a standalone optimisation engine, but as a \emph{heuristic subroutine} embedded within a larger hybrid pipeline. Its role is to explore a structured combinatorial search space and generate candidate solutions that can be refined, validated, or combined using classical methods.

\medskip 

\paragraph{Finance: portfolio selection and risk-aware allocation.}
Portfolio optimisation provides a natural example of a discrete optimisation problem that can be mapped to QUBO form. The task is to allocate capital across a set of assets so as to balance expected return against risk, subject to constraints such as budget limits, cardinality restrictions, or transaction costs. These constraints can be encoded as quadratic penalties, leading to a QUBO formulation.

Quantum annealers have been explored as heuristic solvers for such problems, typically within hybrid workflows in which classical algorithms handle data processing and post-selection, while the annealer searches the combinatorial space of possible portfolios~\cite{ORUS2019100028}. In this setting, the value of QA lies in its ability to rapidly generate diverse candidate solutions rather than to guarantee optimality.

\medskip 

\paragraph{Energy systems and unit commitment.}
Power-grid optimisation involves determining which generators to activate and how to distribute load in order to meet demand while respecting operational constraints. These problems involve large numbers of binary decisions and dense coupling between variables, making direct implementation on current annealing hardware challenging.

As a result, practical approaches typically rely on hybrid strategies in which the global optimisation problem is decomposed into smaller subproblems that can be embedded on the hardware, while classical routines coordinate the overall solution process~\cite{Quinton:2024rkl}. In this context, QA serves as a specialised tool for exploring difficult subspaces of the combinatorial search landscape.

\medskip 

\paragraph{Biology, chemistry, and scientific modelling.}
Many problems in molecular science can be viewed as searches for low-energy configurations of complex systems. While state-of-the-art classical methods (such as deep-learning approaches for protein structure prediction) dominate large-scale applications, QA has been explored as a heuristic tool for simplified or coarse-grained discrete subproblems. These include lattice-based folding models and combinatorial selection tasks that can be expressed in QUBO form~\cite{PhysRevResearch.4.043013}.

Similar ideas appear in other areas of scientific modelling, such as parameter estimation from indirect data. For example, groundwater modelling problems have been mapped to QUBO form by discretising spatial parameters and minimising the mismatch between simulated and observed measurements~\cite{o2018approach}. In such cases, QA acts as a search heuristic within a broader modelling and inference workflow.

\medskip 

Across these examples, a consistent pattern emerges. Quantum annealers are most naturally applied when a larger computational task contains a \emph{core discrete optimisation subproblem} of moderate size that can be isolated and mapped to an Ising or QUBO model. The surrounding tasks---data preparation, constraint handling, decomposition, and verification---remain classical.

This leads to a clear picture of the most promising near-term role of QA in industrial settings. Rather than replacing classical optimisation algorithms, QA is best understood as a \emph{specialised component in hybrid workflows}, particularly in regimes where:
\begin{itemize}
\setlength\itemsep{2pt}
\item high-quality approximate solutions are sufficient,
\item the search space is structured but difficult to explore classically, and
\item multiple candidate solutions can be exploited downstream.
\end{itemize}

In this sense, the practical value of QA lies less in asymptotic speedup and more in its ability to provide a complementary search and sampling primitive within complex optimisation pipelines.

\subsection{Summary: when and why QA is useful}
\label{subsec:applications_synthesis}

\begin{table*}[ht]
\centering
\caption{Summary of major application domains of QA, organised by the primary role played by the device, the dominant dynamical regime, typical performance metrics, and current level of maturity. Different applications probe different aspects of QA and therefore require distinct criteria for evaluation.}
\label{tab:qa_applications}
 \includegraphics[width=0.93\textwidth]{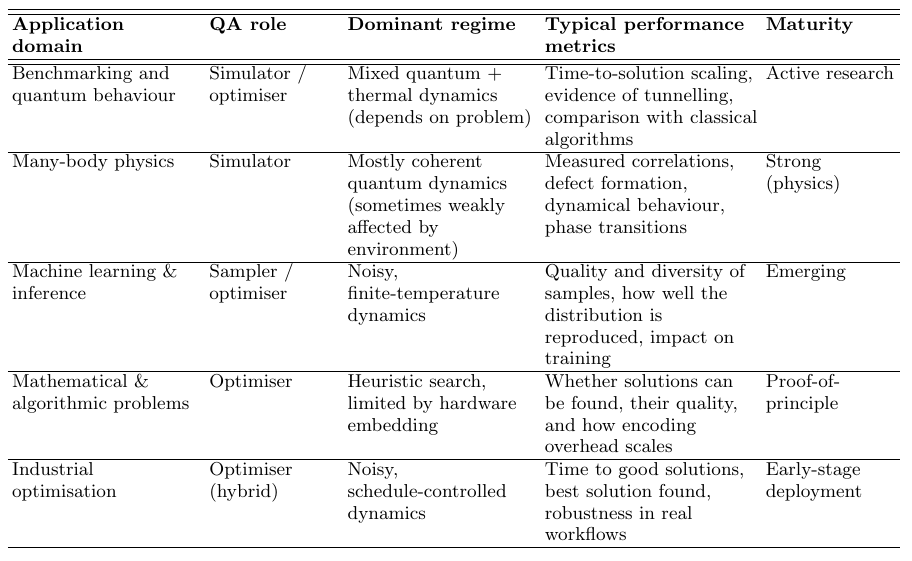}
\end{table*}

Table~\ref{tab:qa_applications} summarises a central message of this section: quantum annealers do not correspond to a single computational paradigm, but rather to a \emph{family of uses} that probe different aspects of the same physical system. The apparent diversity of applications can be understood through three primary roles---optimiser, sampler, and simulator---but the crucial point is that each role comes with its own notion of performance, its own relevant observables, and its own regime of validity.

This perspective helps resolve a common source of confusion in the literature. Discussions of ``quantum advantage'' are often framed as if QA were a single algorithm with a single objective. In practice, however, optimisation, sampling, and simulation are fundamentally different tasks. A device that does not outperform classical methods as an optimiser may still provide useful sampling behaviour, or access to dynamical regimes that are difficult to reproduce classically. Evaluating QA therefore requires matching the \emph{task} to the \emph{metric}, rather than seeking a universal notion of advantage.

A more productive question is therefore not whether QA is generically superior, but \emph{when its physical dynamics align with the structure of the problem}. Across the applications discussed above, three recurring features emerge:

\begin{itemize}
\setlength\itemsep{2pt}
\item \emph{Structured energy landscapes:} QA is most effective when the problem can be expressed as an Ising or QUBO model whose landscape contains features such as barriers, degeneracies, or clustered minima that can be explored efficiently by a combination of tunnelling, relaxation, and non-adiabatic dynamics.

\item \emph{Tolerance to approximation:} many successful applications do not require exact ground states, but rather high-quality approximate solutions or representative samples drawn from a low-energy distribution.

\item \emph{Hybrid decomposition:} in practical settings, QA is typically embedded within a larger workflow in which classical algorithms handle decomposition, constraint management, and refinement, while the annealer is used to explore a well-defined combinatorial core.
\end{itemize}

These features point to a consistent interpretation of QA as a \emph{specialised dynamical primitive} rather than a general-purpose optimiser. Its value lies in providing a physically distinct mechanism for exploring complex energy landscapes, one that can complement, rather than replace, classical approaches.

From this viewpoint, the most persuasive near-term applications of QA are not those that aim to demonstrate a broad asymptotic speedup across generic optimisation problems, but those in which the physical dynamics of the device are well matched to the structure of the task. In optimisation, this means identifying structured subproblems for which heuristic exploration can generate useful candidate solutions. In sampling, it means exploiting the nontrivial low-energy distributions produced by the device. In physics, it means using annealers as controllable platforms for studying many-body dynamics in regimes that are difficult to access classically.

Taken together, these applications suggest that the significance of QA lies less in any single headline claim than in the emergence of a distinctive computational and physical toolkit. Annealers are most valuable when they are used in ways that respect the strengths and limitations of present-day hardware: moderate-sized encodable problems, approximate or stochastic objectives, and workflows in which quantum dynamics complements rather than replaces classical processing. This perspective provides a natural starting point for assessing the challenges and opportunities that will shape the next stage of the field.

\section{Future perspectives and concluding remarks}
\label{sec:future}

QA occupies a distinctive position within the broader landscape of quantum technologies. It is neither simply an analogue optimiser nor merely a restricted form of adiabatic quantum computation. Rather, it is a programmable framework in which optimisation, sampling, and many-body dynamics are realised through the controlled evolution of interacting quantum systems. As the preceding sections have shown, these perspectives are not separate: the same device can function as a heuristic optimiser on one task, a non-equilibrium sampler on another, and a laboratory for quantum dynamics on a third.

This breadth makes the future of QA both promising and difficult to assess. Progress is unlikely to be captured by a single benchmark or a single notion of advantage. Instead, it will depend on how successfully the field aligns three coupled axes: \emph{hardware fidelity}, \emph{dynamical control}, and \emph{problem structure}. Improvements in coherence, connectivity, calibration, and control precision determine which Hamiltonians can be implemented reliably. Advances in schedules, driver design, error suppression, and hybrid workflows determine how those Hamiltonians are explored. Equally, progress in encoding and benchmarking determines which computational or physical tasks are genuinely well matched to annealing dynamics.

From this perspective, the central challenge is not simply to build larger devices, nor simply to search for isolated speedups, but to understand when the physical dynamics of annealing provide a useful computational or scientific resource. This requires clearer distinctions between exact optimisation, approximate optimisation, sampling, and simulation, as well as fair comparisons with rapidly improving classical methods.

In the remainder of this section we highlight several directions that appear particularly important for the next stage of the field: mitigating noise and control errors, reshaping spectral bottlenecks through richer Hamiltonians, developing more effective algorithmic workflows on fixed hardware, and clarifying the boundary between classically tractable and genuinely distinctive annealing regimes.

\subsection{Hardware limitations, error mitigation, and the challenge of fault tolerance}

Among the three axes identified in the opening of this section---hardware fidelity, dynamical control, and problem structure---hardware fidelity remains the most immediate constraint on present-day quantum annealers. Unlike idealised models of adiabatic quantum computation, experimental devices operate as open systems subject to noise, control imperfections, and finite temperature. As a result, the realised dynamics deviates from the intended Hamiltonian evolution, and the output of the computation is inherently stochastic.

Several sources of imperfection play a dominant role in current platforms. Environmental coupling induces relaxation and dephasing, disrupting coherent evolution and enabling unwanted transitions between energy levels. Control errors and finite precision in programmable parameters mean that the implemented Hamiltonian may differ systematically from the intended problem instance. In addition, finite temperature leads to thermal excitation of higher-energy states during the anneal. These effects collectively limit the probability of reaching the ground state and, more generally, distort the low-energy distribution sampled by the device.

These limitations highlight a fundamental distinction between QA and the circuit model of quantum computation. In gate-based architectures, the theory of quantum error correction establishes that, below a threshold noise rate, arbitrarily long computations can in principle be performed reliably using fault-tolerant protocols. No analogous general fault-tolerance theorem is currently known for QA. While partial protection schemes can be constructed, it remains an open question whether fully scalable fault tolerance can be achieved within the annealing paradigm.

As a consequence, the central near-term strategy in QA is not full error correction, but \emph{error suppression and mitigation}. Rather than attempting to eliminate noise entirely, current approaches aim to reduce its impact on physically realistic computations and to extract reliable information from imperfect devices. This shift in emphasis reflects both practical constraints and the hybrid, approximate nature of many QA applications.

A widely studied class of suppression techniques is \emph{quantum annealing correction} (QAC), in which each logical variable is encoded into multiple physical qubits coupled by strong penalty terms that energetically penalise inconsistencies. The logical state is recovered through a decoding step, typically majority voting. Originally proposed as an analogue of repetition codes for annealing, QAC has been demonstrated experimentally to mitigate local errors such as single-qubit flips and to reduce the impact of broken chains arising from embedding~\cite{Jordan:2006gzz,Pudenz:2013mwj}. 

A systematic extension of this idea is \emph{nested} quantum annealing correction (NQAC), in which each logical qubit is replaced by a larger encoded block~\cite{Vinci:2015hhj}. Increasing redundancy effectively amplifies the energy scale of the problem Hamiltonian, making thermal excitations and control errors less likely to disrupt the encoded configuration. This behaviour is often described heuristically as a reduction in effective temperature. The trade-off is clear: allocating more physical qubits per logical variable improves robustness but reduces the size of problems that can be embedded on a fixed device. Identifying the optimal balance between protection and problem size is therefore a key practical consideration.

Complementary to redundancy-based schemes are approaches grouped under \emph{error mitigation}. These methods do not attempt to protect the computation directly, but instead aim to estimate or reduce the bias introduced by noise in measured observables. A prominent example in the circuit model is \emph{zero-noise extrapolation} (ZNE), where experiments are repeated at different effective noise levels and extrapolated to the zero-noise limit~\cite{PhysRevLett.119.180509}. In QA, implementing analogous ideas is more challenging due to continuous-time open-system dynamics, but the underlying principle remains: by varying controllable parameters that influence noise while maintaining a well-characterised Hamiltonian, one can combine measurements to infer less noisy results.

Error mitigation is particularly relevant when annealers are used as samplers or simulators rather than strict optimisers. In these regimes, the objective is often to estimate expectation values, correlations, or distributional properties, and even modest reductions in systematic bias can significantly improve the usefulness of the results. 

These approaches point to a pragmatic near-term paradigm. Improvements in hardware fidelity, through better calibration, reduced noise, and enhanced control precision, remain essential, but they are complemented by algorithmic strategies that adapt to imperfect devices. Rather than relying on a single universal solution, current practice combines moderate redundancy, careful parameter tuning, and problem-specific mitigation techniques. Within the broader three-axis framework, this subsection highlights that advances in hardware fidelity must be accompanied by compatible strategies in dynamical control and problem design if QA is to deliver reliable and scalable performance.

\subsection{Mitigating spectral bottlenecks: catalysts and richer drivers}
\label{subsec:catalysts}

Whereas the previous subsection focused on limitations arising from imperfect hardware fidelity, an independent and more fundamental challenge for QA lies in the \emph{spectral structure} of the annealing Hamiltonian. This issue pertains primarily to the axis of \emph{dynamical control}, and persists even in the idealised limit of noiseless evolution.

As discussed earlier, the runtime required to follow the instantaneous ground state is governed by the minimum energy gap between the ground and first excited states encountered during the anneal. For certain classes of optimisation problems, this gap can become extremely small as system size increases. In particular, instances exhibiting first-order quantum phase transitions along the annealing path are associated with exponentially small gaps, rendering naive adiabatic schedules impractically slow. From this perspective, small gaps represent an intrinsic dynamical bottleneck, independent of noise or control errors.

A central line of research therefore seeks not only to optimise the annealing \emph{schedule}, but to modify the annealing \emph{path} itself in order to reshape the spectrum. A broad class of proposals introduces additional intermediate terms into the Hamiltonian,
\begin{equation}
H(t)=A(t)H_{\mathrm{driver}}+B(t)H_{\mathrm{problem}}+g(t)\,H_{\mathrm{cat}},
\end{equation}
where the auxiliary term $H_{\mathrm{cat}}$ vanishes at the beginning and end of the evolution, $g(0)=g(\tau)=0$. Because such terms alter the intermediate dynamics without changing the initial and final Hamiltonians, they are referred to as \emph{catalysts}.

In simplified theoretical settings, catalysts can have striking effects. For example, in the well-studied ferromagnetic $p$-spin model, the addition of suitably chosen transverse interactions can convert a first-order phase transition into a second-order one, thereby replacing an exponentially small minimum gap with a polynomially closing gap \cite{PhysRevE.85.051112,Seoane_2012}. This illustrates a key principle: \emph{the computational difficulty of an annealing process is not determined solely by the problem Hamiltonian, but by the entire interpolation path in Hamiltonian space}. Modifying that path can fundamentally change the dynamical bottlenecks encountered during the evolution.

At the same time, several important limitations must be emphasised. The effectiveness of a given catalyst is highly problem-dependent: interactions that improve performance for one class of instances may have little effect, or even degrade performance, for others. Explicit counterexamples are known in which non-stoquastic or transverse couplings fail to increase the minimum spectral gap \cite{Albash2019,Takada_2021}. Moreover, current hardware implementations typically support only a restricted class of driver Hamiltonians, most commonly a uniform transverse field combined with programmable pairwise $ZZ$ couplings. As a result, many catalyst constructions remain theoretical or have been explored only in small-scale simulations.

Despite these caveats, this line of work is valuable for two complementary reasons. First, it provides a framework for understanding which features of an energy landscape give rise to severe spectral bottlenecks, thereby linking problem structure to dynamical complexity. Second, it motivates the development of more expressive annealing hardware, including programmable transverse couplings, spatially varying drivers, and effective non-stoquastic interactions. Recent proposals suggest that some of these effects may be approximated even on constrained devices through auxiliary constructions or schedule engineering \cite{Banks:2025efv}.

More broadly, the study of catalysts highlights a shift in perspective: improving QA is not only a matter of running slower schedules on fixed Hamiltonians, but of \emph{designing the Hamiltonian path itself}. This viewpoint connects directly to the dynamical-control axis identified above, and suggests that future progress will depend as much on expanding the space of accessible Hamiltonians as on increasing hardware scale or coherence.
\subsection{Algorithmic directions: schedules, reverse annealing, and hybrid workflows}
\label{subsec:alg_future}

Beyond hardware improvements and Hamiltonian design, a substantial degree of flexibility in QA arises at the level of \emph{algorithmic control}. This flexibility lies primarily along the axis of \emph{dynamical control}, and reflects the fact that even with a fixed problem Hamiltonian, the time-dependent evolution can be engineered in ways that significantly alter performance.

A central ingredient is the annealing \emph{schedule}, which determines how the driver and problem Hamiltonians are interpolated in time. Departures from simple monotonic schedules---such as the introduction of pauses, quenches, or non-monotonic paths---can modify the interplay between coherent evolution and thermal relaxation~\cite{perdomo2011study, marshall2019power, ohkuwa2018reverse}. For example, pauses inserted near regions of small spectral gap can enhance thermal relaxation into low-energy states, while rapid changes in the schedule can induce non-adiabatic transitions that help the system escape local minima. More generally, schedule design provides a practical means of shaping the \emph{dynamical trajectory} of the system through its energy landscape without altering the underlying Hamiltonian.

An even more direct departure from the standard annealing paradigm is provided by \emph{reverse annealing}. In this protocol, the computation is initialised not in a trivial ground state, but in a classical candidate solution. The system is then partially driven back toward the transverse-field regime and subsequently re-annealed toward the problem Hamiltonian. This procedure effectively implements a form of local search, in which quantum fluctuations and thermal effects enable exploration of configurations in the neighbourhood of the initial state. From an algorithmic perspective, reverse annealing transforms QA from a global search heuristic into a controllable refinement tool that can be repeatedly applied to improve candidate solutions.

These ideas naturally lead to a broader paradigm in which quantum annealers are embedded within \emph{hybrid quantum--classical workflows}~\cite{venturelli2015quantum}. In this setting, classical algorithms are responsible for tasks such as problem decomposition, candidate generation, and constraint handling, while the annealer acts as a specialised subroutine for exploring complex energy landscapes. A typical workflow alternates between classical and quantum steps: classical routines identify promising regions of the search space, the annealer generates low-energy samples or performs local exploration, and classical post-processing refines and evaluates the resulting configurations. This iterative structure reflects the empirical behaviour of current devices, which are most naturally viewed as generators of \emph{ensembles} of low-energy states rather than as deterministic optimisers.

This shift in algorithmic perspective has important implications for benchmarking and performance evaluation. Rather than focusing exclusively on asymptotic speedups for exact optimisation, it becomes more appropriate to consider metrics that capture the practical utility of the annealer within a workflow. Examples include \emph{time-to-good-solution}, the quality of the best solution found within a fixed computational budget, and the diversity or usefulness of the returned sample set. These criteria align with the approximate optimisation and sampling-oriented viewpoints developed earlier in the paper, and emphasise that the value of QA is often realised through its integration into larger computational pipelines rather than through standalone execution.

More broadly, these developments point to a re-interpretation of QA as a \emph{programmable dynamical primitive}. Its effectiveness depends not only on the problem Hamiltonian, but on how the evolution is orchestrated in time and how it is embedded within a wider algorithmic context. Progress along this direction is therefore likely to come from co-design across schedules, problem structure, and classical control logic, rather than from improvements in any single component in isolation.

\subsection{Classical simulation and tensor-network viewpoints}
\label{subsec:TN}

An important perspective on the future of QA comes not from hardware or algorithm design alone, but from progress in \emph{classical simulation}. This direction primarily probes the axis of \emph{problem structure}, while also constraining the regimes of \emph{dynamical evolution} that remain classically tractable. In particular, advances in tensor-network and related variational methods have substantially reshaped the boundary between problems that require quantum resources and those that can be simulated efficiently on classical hardware.

Tensor-network approaches~\cite{orus2014practical, schollwock2011density} represent quantum states using structured decompositions that exploit the fact that many physically relevant systems exhibit only limited quantum correlations (entanglement). In practice, this means that the full complexity of the quantum state can often be approximated accurately by retaining only the dominant correlations in the state. Different tensor-network architectures implement this idea in different ways, for example through hierarchical constructions such as \emph{tree tensor networks} (TTNs)~\cite{Shi:2006zz} or by systematically organising correlations across length scales, as in the \emph{multi-scale entanglement renormalisation ansatz} (MERA)~\cite{Vidal:2007hda,Vidal:2008zz,Levin:2006jai,Evenbly:2015uca}. Related developments include neural-network quantum states and other variational representations~\cite{carleo2017solving}. Although these methods are not efficient in all regimes, they have substantially expanded the range of quantum many-body systems that can be simulated on classical computers with controlled accuracy.

One immediate implication is the availability of increasingly powerful \emph{benchmarking baselines}. Whenever classical simulation is feasible at relevant system sizes, it provides a controlled reference for validating annealing dynamics. Comparisons between experimental data and tensor-network simulations can reveal whether a device follows the intended Hamiltonian evolution and help disentangle the roles of coherent dynamics, thermal effects, and control errors. This is particularly relevant for QA implementations with quasi-one-dimensional structure, such as domain-wall encodings or chain embeddings, where matrix-product-state or matrix-product-operator methods often provide efficient and accurate classical descriptions~\cite{crosson2016simulated}. More generally, many annealing Hamiltonians fall into the class of \emph{stoquastic} models, for which classical simulation is believed to be comparatively easier than for generic quantum systems, though not necessarily efficient in all regimes~\cite{bravyi2017complexity}.

Beyond benchmarking, tensor-network methods provide a complementary role as a \emph{design and diagnostic language}. Many QUBO and Ising encodings rely on penalty constructions that define a low-energy manifold of valid configurations. Tensor-network representations can make the structure of this manifold explicit, revealing when an encoding produces well-separated minima and when it instead leads to near-degeneracies or glassy landscapes associated with small spectral gaps. In this way, classical simulation tools can inform the design of problem instances, driver Hamiltonians, and annealing schedules, helping to identify regimes in which quantum dynamics is most likely to offer an advantage.

At the same time, tensor-network methods have well-understood limitations. Their efficiency depends on restricted entanglement growth and favourable geometry. Highly connected interaction graphs, higher-dimensional structures, or dynamics near quantum critical points can lead to rapid entanglement growth, making accurate classical simulation increasingly costly. From this perspective, the most informative question is not whether a given annealing problem can be simulated classically, but \emph{where the boundary of classical tractability lies}. 

This boundary provides a useful lens for assessing the potential role of QA. Problems that remain well described by low-entanglement tensor-network states are unlikely to exhibit a strong quantum advantage, whereas instances that generate complex entanglement or resist efficient classical representation mark regimes in which annealers may offer distinctive capabilities. In this sense, classical simulation is not merely a competing approach, but an essential tool for mapping the landscape in which QA operates and for identifying the regions where it may provide genuine computational or physical insight.

\subsection{Concluding remarks}

Quantum annealing is most usefully understood not as an idealised algorithm, but as a \emph{physical process}: programmable many-body quantum dynamics applied to energy landscapes that encode computational problems. In the adiabatic limit this framework connects to universal quantum computation. In practice, however, present-day devices operate in an intermediate regime where coherent tunnelling, non-adiabatic transitions, thermal relaxation, and environmental noise all contribute to the evolution. The resulting behaviour is therefore inherently stochastic and strongly instance-dependent, and its evaluation requires benchmarking methodologies that reflect this physical reality.

This perspective leads to a clear reinterpretation of the role of QA. It is neither a universal solver for NP-hard optimisation problems nor a classical heuristic in disguise. Rather, it constitutes a distinct \emph{computational primitive} whose capabilities arise from the interplay of quantum dynamics, thermal effects, and programmable structure. Its most promising uses follow directly from this character: as a generator of structured low-energy samples, as a mechanism for exploring complex energy landscapes through controlled dynamics, as a component within hybrid quantum--classical optimisation pipelines, and as a platform for probing non-equilibrium many-body physics at scale.

A central theme that emerges across these domains is that performance is determined not by any single factor, but by the alignment of three elements: hardware capabilities, dynamical control, and problem structure. Improvements in coherence, connectivity, and calibration expand the accessible space of embeddings; advances in schedule design, reverse annealing, and hybrid workflows refine how that space is explored; and careful formulation of optimisation or sampling tasks determines whether the resulting dynamics are advantageous. Understanding and exploiting this interplay is likely to be more important than seeking broad, instance-independent speedups.

From this viewpoint, the most meaningful notion of progress is not the demonstration of asymptotic advantage in isolation, but the identification of regimes in which QA provides capabilities that are difficult to reproduce classically. These include situations where low-energy sampling is intrinsically valuable, where structured landscapes benefit from dynamical exploration, and where classical simulation becomes limited by entanglement growth or combinatorial complexity. Clarifying the boundaries of these regimes---through improved benchmarking, closer integration with classical methods, and deeper theoretical understanding---remains a central challenge for the field.

Looking forward, continued advances in hardware, control, and algorithmic design will further expand these opportunities. At the same time, progress in classical simulation and optimisation will continue to sharpen the benchmarks against which quantum annealers are assessed. The long-term impact of QA will therefore depend not only on improving devices, but on developing a precise understanding of where they fit within the broader computational landscape.

Beyond its computational interpretation, QA also invites a broader physical perspective. It is natural to ask whether processes analogous to quantum annealing occur in physical systems without external control. In many-body systems, relaxation toward low-energy configurations often proceeds through a combination of quantum tunnelling, thermal activation, and dissipation---mechanisms that closely resemble those exploited in QA. While such processes are not typically programmable or optimised for computation, this analogy suggests that quantum annealing can be viewed as a controlled realisation of broader physical relaxation dynamics. From this perspective, QA not only provides a computational strategy, but also offers a framework for understanding how complex systems explore energy landscapes in nature.

In this sense, QA has already achieved something significant. It has established a concrete and scalable setting in which quantum many-body dynamics can be programmed, measured, and harnessed for computational tasks. Whether its ultimate role lies primarily in optimisation, sampling, or physics, its continued development will contribute both to practical algorithmic techniques and to a deeper understanding of how quantum dynamics can be used as a resource for computation.

\appendix
\section{Expressivity of the Ising framework and computational complexity}
\label{sec:ComplexityTheory}

Quantum annealers are designed to tackle optimisation problems that can be expressed as the task of finding the ground state of an Ising Hamiltonian. As discussed in the main text, this formulation already captures a wide range of problems arising in physics, computer science, and industry. In this appendix we take a broader perspective and ask how expressive this framework is: how large is the class of problems that can be mapped onto Ising models, and what can be said about the intrinsic difficulty of solving them?

Addressing these questions requires concepts from computational complexity theory, which classifies problems according to how the resources required to solve them scale with the problem size. In complexity theory one usually focuses on \emph{decision problems}, which are problems whose answer is simply ``yes'' or ``no''. Many optimisation problems can be converted into equivalent decision problems by asking, for example, whether there exists a configuration whose cost is below a given threshold. Studying decision problems allows different computational tasks to be compared in a unified framework.

The class \textbf{P} consists of decision problems that can be solved exactly by an algorithm whose running time grows at most polynomially with the size of the input. Problems in P are therefore considered ``efficiently solvable'' in practice.
The broader class \textbf{NP}, which stands for \emph{nondeterministic polynomial time}, consists of decision problems with the property that, if a candidate solution is provided, its correctness can be \emph{verified} in polynomial time. In other words, while finding a solution to an NP problem may require an exhaustive search, checking whether a proposed solution is correct can be done efficiently. Every problem in P is also in NP, since any problem that can be solved efficiently can also be verified efficiently. Conversely, whether every problem in NP can also be solved efficiently (that is, whether $\text{P}=\text{NP}$) remains one of the most important open questions in computer science, and is widely believed to have a negative answer.

Without entering too deeply into this question, it is useful to introduce the notion of \emph{NP-completeness}. A decision problem is NP-complete if every other problem in NP can be transformed into it using a polynomial-time reduction. NP-complete problems are therefore among the hardest problems in NP: an efficient algorithm for any one of them would imply efficient algorithms for all NP problems.

Remarkably, the decision version of the Ising ground-state problem (e.g.~asking whether the ground-state energy of a given Ising Hamiltonian is below a specified threshold) is NP-complete. This means that any decision problem in NP can be mapped, in polynomial time, onto an instance of an Ising model. As a consequence, the Ising framework provides a universal \emph{encoding} for NP decision problems. Well-known examples that admit such mappings include \textsc{Subset Sum}, \textsc{Clique}, \textsc{Graph Colouring}, \textsc{Vertex Cover}, \textsc{Hamiltonian Cycle}, and \textsc{3-SAT}. The fact that problems originating in logic, combinatorics, and number theory can all be expressed in this common language underlines the broad theoretical reach of Ising-based optimisation.

It is important to emphasise what this result does \emph{not} imply. The NP-completeness of the Ising ground-state problem establishes that the framework is expressively universal for NP decision problems, but it does not guarantee that such problems can be solved efficiently by QA. Rather, it shows that QA targets a class of problems that are believed to be intrinsically hard for both classical and quantum algorithms in the worst case. The practical question is therefore not one of worst-case complexity, but of typical-case performance, the structure of problem instances, and the role of physical dynamics in exploring complex energy landscapes.

Many optimisation problems of practical interest are not decision problems but are instead formulated as minimisation tasks. These are commonly referred to as \emph{NP-hard} problems. Finding the ground state of a general Ising Hamiltonian is itself NP-hard, reflecting the intrinsic difficulty of the optimisation task. While NP-hardness alone does not guarantee that every NP-hard problem can be mapped onto an Ising model, extensive work has shown that a large number of important optimisation problems do admit such mappings. In particular, explicit constructions have been given for problems such as vertex cover, maximum matching, set cover, knapsack, and job sequencing, among many others~\cite{Lucas:2013ahy}.

These results show that the Ising framework is not merely a convenient representation for a narrow class of toy problems, but a highly expressive language capable of capturing a vast family of computationally hard tasks. This universality is one of the central motivations for QA: a single physical platform can, in principle, encode a wide range of optimisation problems within a common dynamical framework. At the same time, the complexity-theoretic perspective makes clear that expressivity does not imply efficiency, and that the performance of QA must ultimately be assessed in terms of physical dynamics, instance structure, and application-specific criteria.
\section{Reduction of higher-order Hamiltonians to quadratic form}
\label{sec:HigherOrderReduction}

Present-day quantum annealers are designed to implement Hamiltonians that are at most quadratic in binary variables, corresponding to Ising or QUBO models. Many optimisation problems of practical interest, however, contain higher-order interactions involving products of three or more variables. In this appendix we outline a systematic procedure by which such higher-order terms can be reduced to an equivalent quadratic form. The construction is standard in both classical and quantum optimisation and is often referred to as \emph{quadratisation}. More comprehensive surveys and alternative constructions can be found in Ref.~\cite{dattani:2019a}. The presentation here is self-contained and focuses on the essential ideas.

\medskip

{\bfseries General framework.}
Consider a polynomial optimisation problem written in terms of binary variables $\tau_i \in \{0,1\}$,
\begin{equation}
\tilde{H}(\{\tau_i\}) = \sum_{i_1<\cdots<i_k} c_{i_1\ldots i_k}\,\tau_{i_1}\cdots\tau_{i_k} + \text{(lower-order terms)},
\label{eq:general_poly_appendix}
\end{equation}
where $k\ge 3$ for at least some terms. Quantum annealers cannot directly implement monomials of degree greater than two, so the goal is to construct an equivalent Hamiltonian that is quadratic in an enlarged set of variables while preserving the ground-state configurations of the original problem.

The central idea is to introduce auxiliary (or ancilla) binary variables that represent products of the original variables. For example, to eliminate a quadratic product $\tau_i \tau_j$ appearing inside a higher-order term, one introduces an auxiliary variable $\tau_{ij}$ intended to satisfy $\tau_{ij} = \tau_i \tau_j$. This logical relation is then enforced energetically by adding a quadratic penalty Hamiltonian whose minimum is attained if and only if the constraint is satisfied.

A convenient quadratic penalty function enforcing $\tau_{ij} = \tau_i \tau_j$ is
\begin{equation}
Q(\tau_{ij};\tau_i,\tau_j) = \Lambda\bigl(\tau_i \tau_j - 2\tau_{ij}(\tau_i+\tau_j) + 3\tau_{ij}\bigr),
\label{eq:Q_constraint_appendix}
\end{equation}
with $\Lambda>0$. A direct enumeration of the four possible assignments $(\tau_i,\tau_j)\in\{0,1\}^2$ shows that $Q=0$ if and only if $\tau_{ij}=\tau_i \tau_j$, while any violation of the constraint incurs a strictly positive energy penalty proportional to $\Lambda$. Crucially, $Q$ is quadratic in all variables.

Under this condition, the auxiliary variable $\tau_{ij}$ may replace the product $\tau_i\tau_j$ wherever it appears in $\tilde{H}$. Repeating this procedure iteratively allows any monomial of degree $k$ to be reduced to quadratic form by introducing $k-2$ auxiliary variables. The resulting Hamiltonian has the general structure
\begin{equation}
\tilde{H} \longrightarrow \tilde{H}_{\mathrm{quad}}(\{\tau_i\},\{\tau_{ij}\},\ldots) + \sum Q,
\end{equation}
where the sum runs over all penalty terms enforcing the auxiliary-variable constraints. Provided the penalty strengths are chosen sufficiently large compared with the coefficients in the original Hamiltonian, the ground states of $\tilde{H}_{\mathrm{quad}}$ coincide exactly with those of $\tilde{H}$.

\medskip

{\bfseries Example: reduction of a three-body interaction.}
As a concrete example, consider the cubic interaction written in spin variables $s_i=\pm1$,
\begin{equation}
\tilde{H} = s_1 s_2 s_3.
\end{equation}
Using the standard mapping between spin and binary variables,
$
s_i = 2\tau_i - 1,
$
and discarding additive constants that do not affect the optimisation problem, this Hamiltonian becomes
\begin{equation}
\tilde{H} = 8\tau_1\tau_2\tau_3 - 4(\tau_1\tau_2 + \tau_1\tau_3 + \tau_2\tau_3) + 2(\tau_1 + \tau_2 + \tau_3).
\label{eq:cubic_binary_appendix}
\end{equation}

The only term of degree higher than two is $\tau_1\tau_2\tau_3$. To eliminate it, we introduce an auxiliary variable $\tau_{12}$ intended to represent $\tau_1\tau_2$, together with the quadratic penalty term $Q(\tau_{12};\tau_1,\tau_2)$ defined in Eq.~\eqref{eq:Q_constraint_appendix}. Substituting $\tau_{12}$ for $\tau_1\tau_2$ yields the fully quadratic Hamiltonian
\begin{equation}
\begin{aligned}
H &= 8\tau_{12}\tau_3 - 4\tau_{12} - 4(\tau_1\tau_3 + \tau_2\tau_3) + 2(\tau_1 + \tau_2 + \tau_3)\\
&~~~~+Q(\tau_{12};\tau_1,\tau_2).
\end{aligned}
\label{eq:quadratic_example_appendix}
\end{equation}

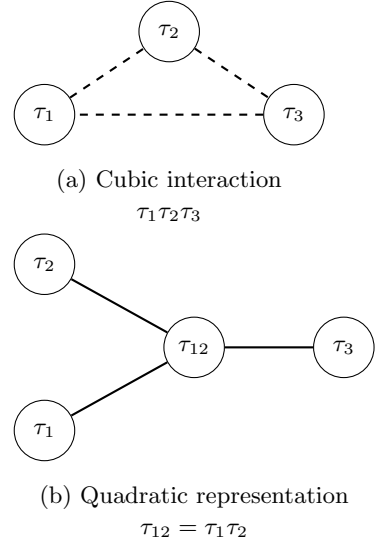
\begin{figure}[t]
\centering
\begin{tikzpicture}[scale=1.1,
    node/.style={circle, draw, minimum size=8mm, inner sep=0pt},
    edge/.style={thick}
]

% -------- Left: original cubic interaction --------
\begin{scope}
\node[node] (t1) at (0,0) {$\tau_1$};
\node[node] (t2) at (1.5,1) {$\tau_2$};
\node[node] (t3) at (3,0) {$\tau_3$};

% triangle (suggesting 3-body)
\draw[edge, dashed] (t1) -- (t2) -- (t3) -- (t1);

\node at (1.5,-.8) {(a) Cubic interaction};
\node at (1.5,-1.2) {$\tau_1\tau_2\tau_3$};
\end{scope}

% -------- Right: quadratized graph --------
\begin{scope}[yshift=-3.8cm]

\node[node] (t1) at (0,0) {$\tau_1$};
\node[node] (t2) at (0,2) {$\tau_2$};
\node[node] (t12) at (1.8,1) {$\tau_{12}$};
\node[node] (t3) at (3.6,1) {$\tau_3$};

\draw[edge] (t1) -- (t12);
\draw[edge] (t2) -- (t12);
\draw[edge] (t12) -- (t3);

\node at (1.8,-.8) {(b) Quadratic representation};
\node at (1.8,-1.2) {$\tau_{12} = \tau_1\tau_2$};

\end{scope}

\end{tikzpicture}
\caption{Reduction of a three-body interaction to quadratic form. (a) A cubic term corresponds to a higher-order interaction among three variables. (b)~Introducing an auxiliary variable $\tau_{12}$ allows the interaction to be expressed using only pairwise couplings, at the cost of additional variables and constraints.}
\label{fig:quadratization}
\end{figure}

For sufficiently large $\Lambda$, any configuration violating the constraint $\tau_{12}=\tau_1\tau_2$ is energetically disfavoured, and the ground states of the quadratic Hamiltonian~\eqref{eq:quadratic_example_appendix} coincide with those of the original cubic interaction. This construction illustrates how higher-order terms can be systematically reduced without introducing non-quadratic interactions elsewhere. The associated connectivity graph is illustrated schematically in Fig.~\ref{fig:quadratization}. The full quadratic Hamiltonian includes additional pairwise couplings arising from the penalty term, but these are omitted in the figure for clarity.

\medskip

{\bfseries Remarks on overhead and connectivity.}
The price paid for this general reduction is an increase in the number of variables and couplings. Each auxiliary variable introduces additional quadratic interactions and enlarges the logical coupling graph. On hardware with restricted native connectivity, representing these interactions may require further embedding overhead, potentially increasing the number of physical qubits needed to realise the problem. 

The central lesson is that the quadratic restriction of present-day annealing hardware is not a fundamental limitation of the framework, but a practical constraint that can be systematically circumvented at the cost of additional resources. In practice, this trade-off can be substantial: quadratisation increases both the size and connectivity of the logical problem, which in turn amplifies embedding overhead and sensitivity to control errors and noise. As a result, the efficiency of a given encoding depends not only on its correctness, but also on how economically it represents higher-order structure within the constraints of the hardware. Designing such encodings is therefore a key component of practical QA.
Nevertheless, such reductions are widely used in both classical and quantum optimisation and greatly expand the range of problems that can be expressed in a quadratic formalism.

%\bibliography{TheLiterature}
%\input{bib.bbl}

\end{document}